\documentclass[journal]{IEEEtran}

\usepackage{cite}
\usepackage[pdftex]{graphicx}
\usepackage[cmex10]{amsmath}
\usepackage{amssymb,amsbsy,upgreek}
\usepackage{microtype,url,enumerate,booktabs,color}
\usepackage{rotating}
\usepackage[T1]{fontenc}

\begin{document}

%\title{Characterizing Sound Field Diffuseness with Spherical Microphone Arrays}
\title{Spherical Harmonic Signal Covariance and Sound Field Diffuseness}

\author{Nicolas Epain and Craig T. Jin
\thanks{CARLab, School of Electrical and Information Engineering, The University of Sydney, NSW, 2006, Australia. Email: craig.jin@sydney.edu.au}}

\maketitle

\begin{abstract}
Characterizing sound field diffuseness has many practical applications, from room acoustics analysis to speech enhancement and sound field reproduction. In this paper we investigate how spherical microphone arrays (SMAs) can be used to characterize diffuseness. Due to their specific geometry, SMAs are particularly well suited for analyzing the spatial properties of sound fields. In particular, the signals recorded by an SMA can be analyzed in the spherical harmonic (SH) domain, which has special and desirable mathematical properties when it comes to analyzing diffuse sound fields. We present a new measure of diffuseness, the COMEDIE diffuseness estimate, which is based on the analysis of the SH signal covariance matrix. This algorithm is suited for the estimation of diffuseness arising either from the presence of multiple sources distributed around the SMA or from the presence of a diffuse noise background. As well, we introduce the concept of a diffuseness profile, which consists in measuring the diffuseness for several SH orders simultaneously. Experimental results indicate that diffuseness profiles better describe the properties of the sound field than a single diffuseness measurement.
\end{abstract}


\IEEEpeerreviewmaketitle

\section{Introduction}
Spherical microphone arrays (SMAs) denote microphone arrangements whereby the sensors are distributed relatively evenly over one or more spherical surfaces, or inside an open sphere. Due to their particular geometry, SMAs provide a panoramic view of the sound waves and thus are particularly well suited for analyzing the sound field in terms of direction of arrival. Over the last decade SMAs have become increasingly popular in the research community: they have been employed for applications as diverse as beamforming~\cite{Meyer2002}, direction-of-arrival estimation~\cite{Khaykin2009}, sound field imaging~\cite{Epain2013}, sound field reproduction~\cite{Betlehem2005,Daniel2004,Moreau2006} and room acoustics analysis~\cite{Gover2002,Gover2004}.

In this work we focus on the specific problem of characterizing sound field \emph{diffuseness} (or diffusivity) using SMAs. Characterizing the diffuseness of the sound field can obviously be of interest to analyze the acoustics of a given room, for instance when designing a concert hall~\cite{Kuttruff2009}. In addition, measuring diffuseness can be seen as a way to estimate the direct-to-reverberant ratio (DRR), which may be used in the context of speech enhancement or beamforming. For instance, a popular speech enhancement technique consists in applying a Wiener filter calculated using a DRR estimate to the output of a beamformer~\cite{McCowan2003}. As well, parametric sound field reproduction and spatial audio coding techniques~\cite{Pulkki2007,Goodwin2008} typically rely on the separation of the sound field into a direct and a diffuse component. This requires estimation of the relative energy of these components and thus an estimate of diffuseness. 

A perfectly diffuse sound field is commonly defined as a sound field with the property that at every location the acoustic energy flows in every direction with equal probability~\cite{Kuttruff2009}. A more practical description arises from the viewpoint of how such a diffuse sound field may be generated. From this viewpoint, a diffuse sound field arises when infinitely many plane waves propagate from every direction in space, with equal strengths and random, mutually uncorrelated phases~\cite{Fahy1985}. In this light, we consider a sound field model in which the sound pressure arises from two components: (i) $Q$ uncorrelated plane waves and (ii) a diffuse noise background:
\begin{equation}
p(\mathbf{r},t) = \sum_{q=1}^Qp_q(\mathbf{r},t) + \sqrt{\nu}\,n(\mathbf{r},t) \ \text{,}
\end{equation}
where $p_q(\mathbf{r},t)$ is the pressure resulting from the $q$-th plane wave, $n(\mathbf{r},t)$ is the pressure resulting from a unit-power diffuse noise background, $\nu$ is the power of the diffuse noise component, $\mathbf{r}$ specifies the location in space and $t$ is time. With this sound field model,  diffuseness increases as either the number of plane waves, $Q$, from different directions increases or as the power, $\nu$, of the diffuse background noise increases. There is thus ambiguity in the sound field model as to the cause of the diffuseness and this ambiguity is real. When there are many plane waves evenly distributed across space, the plane wave component does indeed resemble the diffuse noise background. Acoustic situations with some late reverberation and a few dominant sources or early echoes arise quite frequently. In this work, our aim is to propose a method to characterize diffuseness that: 1) provides a reasonable diffuseness estimate in situations where diffuseness arises from either the presence of multiple uncorrelated sources or the presence of a diffuse noise background; and 2) provides some information regarding how much diffuseness arises from each of these causes.

Different methods have been proposed for the estimation of diffuseness from a set of measured microphone signals. These methods can be divided into four main approaches. One approach consists in measuring the amount of acoustic energy incoming from every direction in space~\cite{Meyer1956,Gover2002,Gover2004}. The core idea behind this approach is that, in a perfectly diffuse sound field, an equal amount of energy is observed for every direction. A second approach consists in measuring the acoustic intensity over time~\cite{Kuttruff2009,Pulkki2007,Abel2006a,Gotz2015}, the idea being that, in a perfectly diffuse sound field, the average acoustic intensity is null. A third approach consists in analyzing the correlations or coherences between the microphone signals~\cite{Jarrett2012,Thompson2012,Schwarz2015,Thiergart2014b}, with the idea being that the directional component of the sound field results in correlated microphone signals, whereas the diffuse component reduces the correlation between the microphone signals. Lastly, the fourth approach consists in measuring the acoustic energy at various points across the sound field~\cite{ISO3741}, with the idea being that the acoustic energy should be constant across space in a diffuse sound field.

In this work, we focus on measuring diffuseness with SMAs. The spherical harmonic sound field representation is a natural framework for SMA signal processing algorithms. Hence, we investigate the specific problem of estimating diffuseness from the spherical harmonic (SH) expansion signals recorded by an SMA. We briefly review previous work using SMAs to measure diffuseness. Pulkki~\cite{Pulkki2007} proposes to use the first four SH signals to estimate the directionality of the sound field as the ratio of the active sound intensity to the acoustic energy density. We show that this method has difficulty characterizing diffuseness that arises from the presence of multiple sources distributed around the SMA. Following a different approach, Gover~\cite{Gover2004} proposes to use SMAs to steer beams in a large number of directions to measure how uniformly the acoustic energy flow is distributed across space. We show that Gover's method does not consider the correlation of signals from different directions. More recently, Jarrett~\cite{Jarrett2012} introduced another algorithm which makes use of the coherence between SH signals to estimate the diffuseness. This technique relies on two assumptions: (i) that only one dominant sound source is present in a given frequency band; and (ii) that an accurate estimate of the dominant source position is available. An advantage of the approach described in this work is that it effectively handles multiple dominant sound sources without requiring an estimate of source positions.

While the formulation of our diffuseness estimator is presented in a previous paper~\cite{Epain2014}, the following are new contributions for this work: 1) We discuss the concept of sound field diffuseness and demonstrate that the eigenvalue spectrum of the SH signal covariance matrix provides significant information regarding diffuseness (Section~\ref{sec:background}); 2) We prove that, in the presence of a single sound source, the diffuseness estimated by our algorithm is equal to the relative noise level (Section~\ref{sec:myMethod}); 3) We provide simulation results comparing our diffuseness estimator to two other algorithms (Section~\ref{sec:myMethod}); and 4) We introduce the concept of a diffuseness profile and provide experimental results showing how diffuseness profiles can be used to characterize sound field diffuseness (Section~\ref{sec:profiles}).

The paper is organized as follows. In Section~\ref{sec:background}, we introduce the concept of diffuseness and describe the properties of SH signals in the presence of a sound field modeled as the sum of a diffuse noise component and a ``directional'' component, which consists of a number of plane waves. In Section~\ref{sec:otherMethods}, we briefly review two existing diffuseness measurement algorithms for SMAs and describe the characteristics of these algorithms. In Section~\ref{sec:myMethod}, we introduce a diffuseness measurement technique and describe how this technique differs from those presented in Section~\ref{sec:otherMethods}. In Section~\ref{sec:profiles} we introduce the concept of a diffuseness profile, which consists in estimating the diffuseness for several SH orders simultaneously, and demonstrate how diffuseness profiles provide additional and new information. Section~\ref{sec:concl} concludes the paper.

%\begin{itemize}
%\item Diffuseness estimation is useful for room analysis, speech enhancement/beamforming, sound field reproduction...
%\item Various approaches have been used. Two types of approaches: energy-based / correlation-based?
%\item SMAs are a new(ish) tool that is particularly well-suited for spatial sound field analysis. Gover proposed a way to adapt Thiele's technique to measure diffuseness. One problem is that the method is based on a purely energetic analysis and certain sound fields may be misinterpreted as diffuse (spherical room). Another problem is that the result will change depending on the SMA used. 
%\item In this paper we propose a method for estimating the diffuseness of a sound field. It is based on the spherical harmonic formalism and thus can be applied to measurements using SMAs. The big advantage of this diffuseness measure is that, in the frequency range of operation of the microphone array, the result is independent of the SMA used. Also it does not suffer from the"spherical room" problem. 
%\item We also introduce the concept of diffuseness profile...
%\item Summary of the different sections
%\end{itemize}

\section{Theoretical Background}
\label{sec:background}
In this section, the concept of sound field diffuseness in the context of the SH sound field representation is briefly surveyed. A simple model for SH signals in the presence of multiple sound sources and diffuse noise is described. Further, the fundamental properties of the SH signal covariance matrix are derived.

\subsection{Signal model}
\label{sec:Model}
Any sound field consisting of incoming waves can be entirely described, in the 2-norm sense, by an infinite series of spherical harmonic functions. In other words, in the frequency domain, the acoustic pressure measured at the spherical coordinates $\mathbf{r}=(r,\theta,\varphi)$ can be written as~\cite{Williams1999}:
\begin{equation}
p(\mathbf{r},k) = \sum_{l=0}^{\infty}\sum_{m=-l}^{l}i^l\, j_l(kr)\,Y_l^m(\theta,\varphi)\,b_{l,m}(k) \ \text{,}
\end{equation}
where $k = 2\pi f c^{-1}$ is the wavenumber, $ j_l(\cdot)$ is the \mbox{order-$l$} spherical Bessel function and $Y_l^m(\theta,\varphi)$ is the \mbox{order-$l$}, \mbox{degree-$m$}, real-valued spherical harmonic function. In this work we use the fully-normalized (N3D) spherical harmonic functions. The \mbox{order-$L$} spherical harmonic expansion of the sound field, which is obtained by keeping the expansion coefficients $b_{l,m}(k)$ up to order~$L$, provides an accurate description of the sound field in the vicinity of the origin -- the higher the order, the greater the area of space inside which this description is accurate.

In the following we employ the time-domain spherical harmonic expansion, that is the time signals $b_{l,m}(t)$ corresponding to the frequency-domain expansion coefficients $b_{l,m}(k)$ via the inverse Fourier transform. We refer to the time-domain, order-$L$, spherical harmonic expansion of the sound field as the \emph{order-$L$ SH signals}. In practice, SH signals are obtained by filtering microphone signals using a set of specially designed digital filters~\cite{Rafaely2005,Moreau2006,Jin2014}. 

In this work we model the SH signals as the sum of two components: 1) the contribution of $Q$ plane-waves propagating from angular directions $\Omega_1$, $\Omega_2$, ..., $\Omega_Q$ where $\Omega_q=(\theta_q,\varphi_q)$; and 2) Gaussian white noise. As we show in the next section, Gaussian white noise provides a reasonable model for a perfectly diffuse noise background because it has similar mathematical properties. In this context the order-$l$, \mbox{degree-$m$} SH signal is given by:
\begin{equation}
b_{l,m}(t) = \sum_{q=1}^Q Y_l^m\!\left(\Omega_q\right) s_q(t)+ \sqrt{\nu}\,n_{l,m}(t) \ \text{,}
\label{eq:Model}
\end{equation}
where $s_q(t)$ is the time signal corresponding to the $q$-th plane-wave, $n_{l,m}(t)$ is a unit-power Gaussian white noise signal, and~$\nu$ is the power of the noise. We further assume that: a) all plane-wave and noise signals are zero-mean; b) all plane-wave and noise signals are mutually uncorrelated; and c) all noise signals have equal power. In other words, we have:
\begin{align}
& \mathrm{E}\left\{s_q(t)\right\}=\mathrm{E}\left\{n_{l,m}(t)\right\}=0 \ \ \forall\, q, l, m \text{ ,} \nonumber \\
& \mathrm{E}\left\{s_q(t)\, s_{q'}(t)\right\} = 0 \ \ \forall\, q'\neq q \text{ ,} \nonumber \\
& \mathrm{E}\left\{s_q(t)\, n_{l,m}(t)\right\} = 0 \ \ \forall\, q, l, m \text{ ,} \nonumber \\
& \mathrm{E}\left\{n_{l,m}(t)\, n_{l',m'}(t)\right\} = 0 \ \ \forall\, (l',m')\neq (l,m) \text{ ,} \nonumber \\
& \mathrm{E}\left\{n_{l,m}^2(t)\right\} = 1 \ \ \forall\, l, m \text{ ,} 
\label{eq:uncorr}
\end{align}
where $\mathrm{E}\left\{\cdot\right\}$ denotes the statistical expectation operator and in practice is often estimated as the time average for the considered period, $t = 1, 2, \dots, T$:
\begin{equation}
\mathrm{E}\left\{s(t)\right\} \approx \frac{1}{T}\sum_{t=1}^{T}s(t) \text{ .}
\end{equation}
According to the signal model, we also define the relative noise level, $\beta$:
\begin{equation}
\beta=\frac{\sum_{l,m}\mathrm{E}\left\{\nu\,n_{l,m}^2(t)\right \}}{\sum_{l,m}\mathrm{E}\left\{{b_{l,m}}^2(t)\right \}} = \frac{(L+1)^2\nu}{\sum_{l,m}\mathrm{E}\left\{{b_{l,m}}^2(t)\right \}} \ \ \text{.}
\end{equation}
As this multi-source model differs from the single-source model that is commonly considered in the literature, the motivation for this multi-source model is made clear in the following section. Note, however, that $\beta$ corresponds to the common defintion of diffuseness in the single-source case.

\subsection{SH Signal Covariance}
\label{subsec:shCov}
In order to characterize the diffuseness of the sound field, we propose to analyze the structure of the SH signal covariance matrix. In the context of the model defined by Equation~\eqref{eq:Model}, the covariance between two different SH signals is given by:
\begin{align}
C_{bb}(l,l',m,m') & = \mathrm{E}\left\{b_{l,m}(t)\,b_{l',m'}(t) \right\} \nonumber \\
& = \mathrm{E}\left\{\left(\sum_{q=1}^Q Y_l^m\!\left(\Omega_q\right) s_q(t)+ \sqrt{\nu}\,n_{l,m}(t)\right)\right. \dots \nonumber \\
& \ \ \ \left.\times \left(\sum_{q=1}^Q Y_{l'}^{m'}\!\left(\Omega_q\right) s_q(t)+ \sqrt{\nu}\,n_{l',m'}(t)\right) \right\}
\end{align}
Using Equation~\eqref{eq:uncorr}, we have:
\begin{align}
C_{bb}(l,l',m,m') & = \sum_{q=1}^Q Y_l^m\!\left(\Omega_q\right)Y_{l'}^{m'}\!\left(\Omega_q\right)\mathrm{E}\left\{s_q^2(t)\right\} \ \dots \nonumber \\
& \hspace{5em} + \delta_{l,l'}\, \delta_{m,m'} \nu
\label{eq:shCovGen}
\end{align}

\begin{figure}[t]
\centerline{ \includegraphics[width=.495\columnwidth]{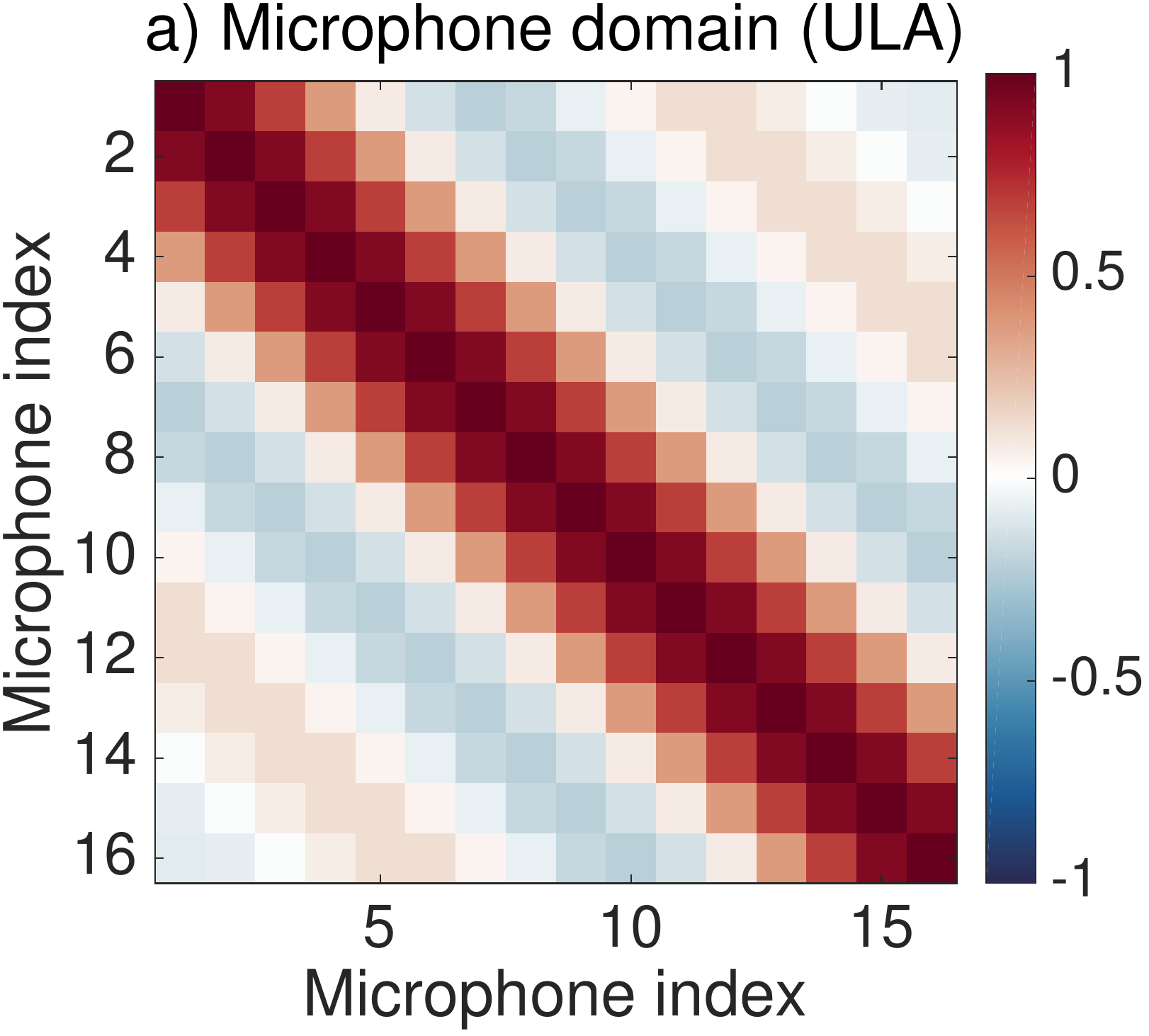}
\includegraphics[width=.49\columnwidth]{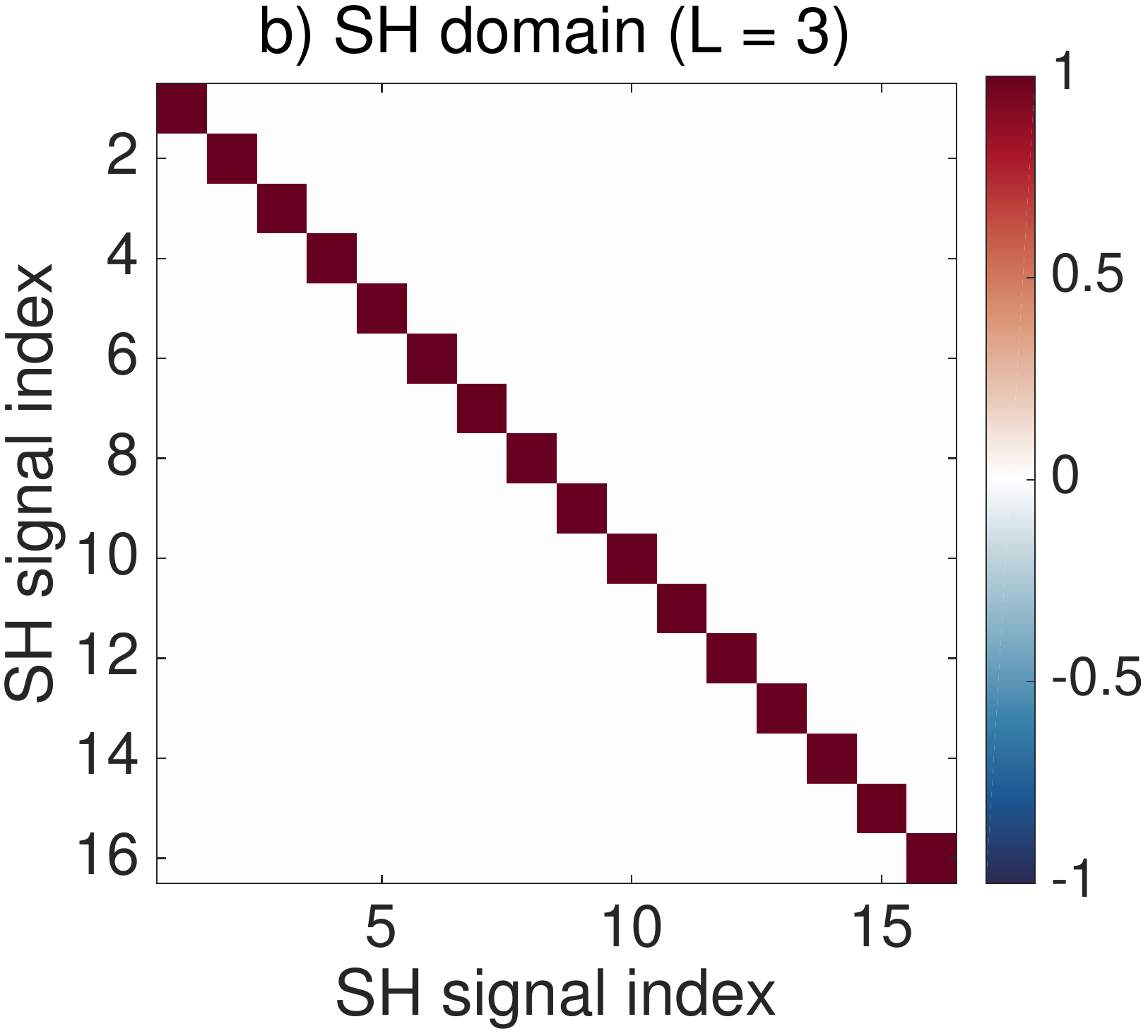} }
\caption{Comparison of the signal covariance matrices observed for a perfectly diffuse sound field in: a) the microphone domain, with a 30~cm long uniform linear array (ULA) consisting of 16~microphones at 2~kHz ; and b) in the spherical harmonic domain, with L=3.}
\label{fig:CovMatricesUlaVsSh}
\end{figure}

We now show that in the presence of a perfectly diffuse sound field, the SH signals are mutually uncorrelated and have equal power. A perfectly diffuse sound field is commonly described as consisting of infinitely many plane waves incoming from every direction in space, with equal strengths and random, mutually uncorrelated phases~\cite{Fahy1985}. The SH signal covariance in this scenario can be calculated by taking the limit of Equation~\eqref{eq:shCovGen} for a continuous distribution of plane waves over the sphere ($Q\rightarrow\infty$), setting $\nu=0$ and replacing the term $\mathrm{E}\left\{s_q^2(t)\right\}$ by a direction-independent power density, $\rho$. We then obtain, using the orthonormality property of the spherical harmonic functions:
\begin{align}
C^{\mathrm{(dif)}}_{bb}(l,l',m,m') & = \int_{\Omega\in S^2}\!\!\!\!Y_l^m\!\left(\Omega\right)Y_{l'}^{m'}\!\left(\Omega\right)\rho\,\mathrm{d}\Omega \nonumber \\
& = \rho \int_{\Omega\in S^2}\!\!\!\!Y_l^m\!\left(\Omega\right)Y_{l'}^{m'}\!\left(\Omega\right)\,\mathrm{d}\Omega \nonumber \\
& = \delta_{l,l'}\, \delta_{m,m'} \rho \ \text{.}
\label{eq:perfDiff}
\end{align}
Therefore, in a perfectly diffuse sound field, SH signals are mutually uncorrelated and have equal power. Note that this result is different from the correlation observed between raw microphone signals. In the presence of a perfectly diffuse sound field, the signal coherence between two omnidirectional microphones separated by distance $\alpha$ is equal to $\mathrm{sin}(k\alpha)/k\alpha$~\cite{Cook1955}. This difference is illustrated in Figure~\ref{fig:CovMatricesUlaVsSh}.

Two important observations should be made at this stage. First, the $\delta_{l,l'}\, \delta_{m,m'} \nu$ term in Equation~\eqref{eq:shCovGen} is similar to the covariance observed for a perfectly diffuse sound field (defined by $Q\rightarrow\infty$ and setting $\nu=0$) as shown in Equation~\eqref{eq:perfDiff}. In other words, the noise in the signal model described by Equation~\eqref{eq:Model} represents a perfectly diffuse noise background, and increasing the relative noise level, $\beta$, increases the diffuseness of the sound field. Second, observe that a highly diffuse sound field can be obtained by distributing many uncorrelated sources over the sphere and that by increasing~$Q$, the number of plane waves in our model, the sound field becomes progressively more and more diffuse. These two different causes of diffuseness (increasing diffuse noise level and increasing the number of plane waves from different directions) constitute a fundamental and real ambiguity associated with the concept of diffuseness. Nevertheless, we show that there is a means to deal with this ambiguity that is related to the signal covariance matrix.

To begin, we provide the following intuition. As the sound field becomes more and more diffuse, one would expect the signal covariance matrix to become more and more similar to the signal covariance matrix for a perfectly diffuse sound field. For a uniform linear array, one expects to obtain the sinc function pattern for the signal covariance matrix as shown in Figure~\ref{fig:CovMatricesUlaVsSh}. Let us consider now a spherical microphone array. In order to illustrate the transition from a direct to a diffuse sound field we calculate the SH signal covariance matrix obtained when more and more uncorrelated plane waves with equal signal power are evenly distributed over the sphere. In this simulation we set $\beta$ to 0 (no diffuse background) and the plane-wave signals are generated as 1024-sample long Gaussian white noise signals. The plane wave source directions form a $Q$-point sphere packing~\cite{SloanePackings}. For each number of plane-waves, $Q$, we calculate the mismatch, $\xi$, between the pattern of the observed SH signal covariance matrix, $\mathbf{C_{bb}}$, and that of the covariance matrix corresponding to a perfectly diffuse sound field, $\rho\,\mathbf{I}_{(L+1)^2}$, as:
\begin{align}
\xi & =\left\|\frac{\mathbf{C_{bb}}}{\left\|\mathbf{C_{bb}}\right\|_2} - \frac{\rho\,\mathbf{I}_{(L+1)^2}}{\left\|\rho\,\mathbf{I}_{(L+1)^2}\right\|_2} \right\|_{\mathrm{F}}^2\bigg/\left\|\frac{\rho\,\mathbf{I}_{(L+1)^2}}{\left\|\rho\,\mathbf{I}_{(L+1)^2}\right\|_2}\right\|_{\mathrm{F}}^{2}\nonumber \\
& = \frac{1}{(L+1)^2}\left\|\frac{\mathbf{C_{bb}}}{\left\|\mathbf{C_{bb}}\right\|_2} - \mathbf{I}_{(L+1)^2} \right\|_{\mathrm{F}}^2\ \text{,}
\end{align}
where $\left\|\cdot\right\|_{\mathrm{F}}$ denotes the Frobenius norm, $\left\|\cdot\right\|_{\mathrm{2}}$ denotes the matrix 2-norm (largest eigenvalue) and $\mathbf{I}_N$ is the $N$-dimensional identity matrix. The mismatch $\xi$ can be described as the energy of the difference between the two matrix patterns relative to the energy of the reference matrix pattern.

\begin{figure}[t]
\includegraphics[width=\columnwidth]{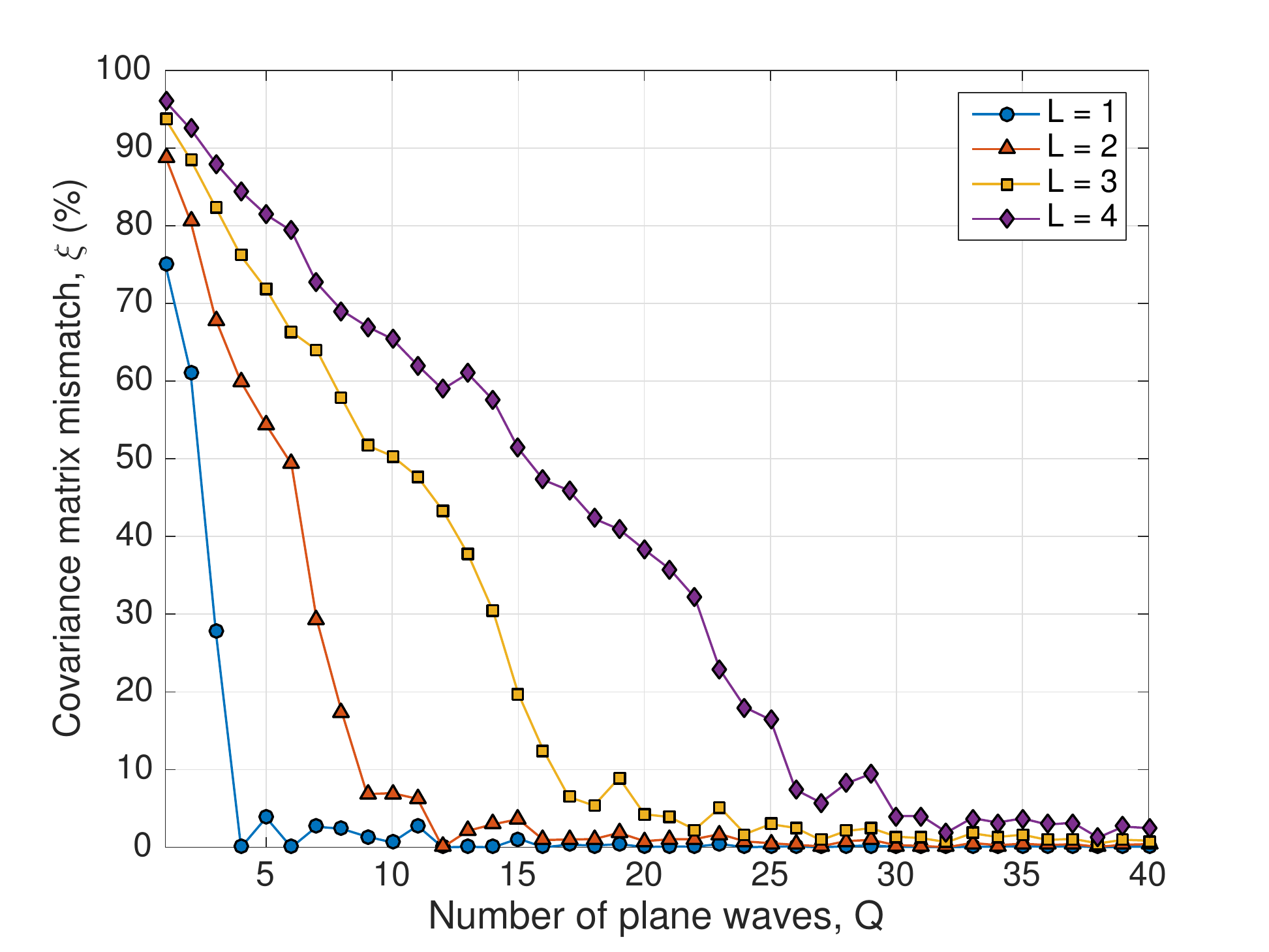}
\caption{This plot illustrates the transition from a direct to a diffuse sound field observed when an increasing number of uncorrelated plane waves are distributed over the sphere.}
\label{fig:DirectDiffuseTransition}
\end{figure}

The results of the simulation are plotted in Figure~\ref{fig:DirectDiffuseTransition}. When there are few plane waves, the SH signal covariance matrix is very dissimilar to that observed in the presence of a perfectly diffuse sound field and $\xi$ is close to 100\%. On the other hand, when the number of plane waves distributed over the sphere increases, the SH signal covariance matrix becomes almost identical to $\rho\,\mathbf{I}_{(L+1)^2}$. In other words, the diffuseness of the sound field increases as more plane waves are distributed over directions and this transition is relatively smooth.

Another significant observation is that, the higher the SH order, $L$, the slower the transition from direct to diffuse. Said differently, this means that it takes more plane waves to make the sound field appear diffuse when considering SH signals up to order~4 than up to order~1. This makes sense because a higher SH order corresponds to a higher angular resolution and this higher resolution makes it possible to resolve sources that would appear as a continuous distribution at lower orders. Lastly, note that it takes approximately $(L+1)^2$ plane waves to make the sound field appear perfectly diffuse at order $L$. Again this stands to reason because, as the sound field consists of~$Q$ plane waves, the rank of the covariance matrix is at most~$Q$. Thus $Q$ has to be greater than the number of SH signals, $(L+1)^2$, for the covariance matrix to be full-rank, as is the case in the presence of a perfectly diffuse sound field.

In summary, in our signal model the diffuseness of the sound field can be varied in two different ways. We can either vary~$\beta$, the relative noise level, or we can vary the position and number of plane waves. We have shown that both a large~$\beta$ and a large $Q$ (with the sources having equal powers and being evenly distributed over the sphere) yielded a perfectly diffuse sound field. However, there is a difference in the nature of the diffuseness induced by these two parameters. To explore this difference further, we consider the eigenvalue spectrum of the signal covariance matrix in the following section.

\subsection{Eigen-decomposition of the SH Signal Covariance Matrix}
\label{subsec:CovMat}
We now demonstrate how the structure of the SH signal covariance matrix, specifically its eigenvalue spectrum, provides significant information relating to the diffuseness of a sound field. Using Equation~\eqref{eq:shCovGen}, the SH signal covariance matrix is given by:
\begin{equation}
\mathbf{C_{bb}} = \mathbf{\Gamma} + \nu\, \mathbf{I}_{(L+1)^2}\ \text{,}
\end{equation}
where 
\begin{align}
& \mathbf{\Gamma} = \sum_{q=1}^Q\mathrm{E}\left\{ s_q^2(t) \right\}\mathbf{y}_q\,\mathbf{y}_q^\mathsf{T} \nonumber \ \text{,} \\
& \mathbf{y}_q=\left[ Y_0^0(\Omega_q),\, Y_1^{-1}(\Omega_q),\, \dots,\, Y_L^L(\Omega_q)\right]^{\mathsf{T}} \ \text{.}
\end{align}
Thus $\mathbf{C_{bb}}$ is the sum of two covariance matrices. The first matrix, $\mathbf{\Gamma}$, which corresponds to the $Q$ plane waves, is a sum of $Q$ rank-one matrices; its rank is at most equal to $Q$ and at most $Q$ of the eigenvalues of $\mathbf{\Gamma}$ are non-zero. The second matrix, which corresponds to the diffuse noise background, is proportional to the identity matrix; its rank is $(L+1)^2$ and its eigenvalues are equal to $\nu$. It follows that the eigenvalues of $\mathbf{C_{bb}}$, $(\sigma_1, \sigma_2, ..., \sigma_{(L+1)^2})$, sorted in decreasing order, are given by:
\begin{equation}
\begin{cases}
\sigma_i=\nu + \omega_i\ & \text{for } 1\leq i\leq Q \\
\sigma_i=\nu & \text{for } Q< i\leq (L+1)^2 \ \text{,} 
\end{cases}
\label{eq:CovSpec}
\end{equation}
where $\omega_i$ denotes the $i$-th eigenvalue of the matrix $\mathbf{\Gamma}$. Note that Equation~\eqref{eq:CovSpec} assumes that $Q$ is less than $(L+1)^2$. If~$Q$ is equal to or greater than $(L+1)^2$, then we have $\sigma_i=\nu+\omega_i\ \ \text{for\ } i=1,\ 2,\ \dots, (L+1)^2$.

\begin{figure}[t]
\begin{tabular}{p{2mm}|cc}
& \textsf{\footnotesize{Covariance matrix}} & \ \ \textsf{\footnotesize{Eigenvalues}} \\
\midrule
\begin{sideways}\ \ \ \ \ \ \ \textsf{\small{$Q = 1$, $\beta=0$}}\end{sideways}
& \includegraphics[trim = 0mm .7mm 0mm 0mm, clip=true,width=.45\columnwidth]{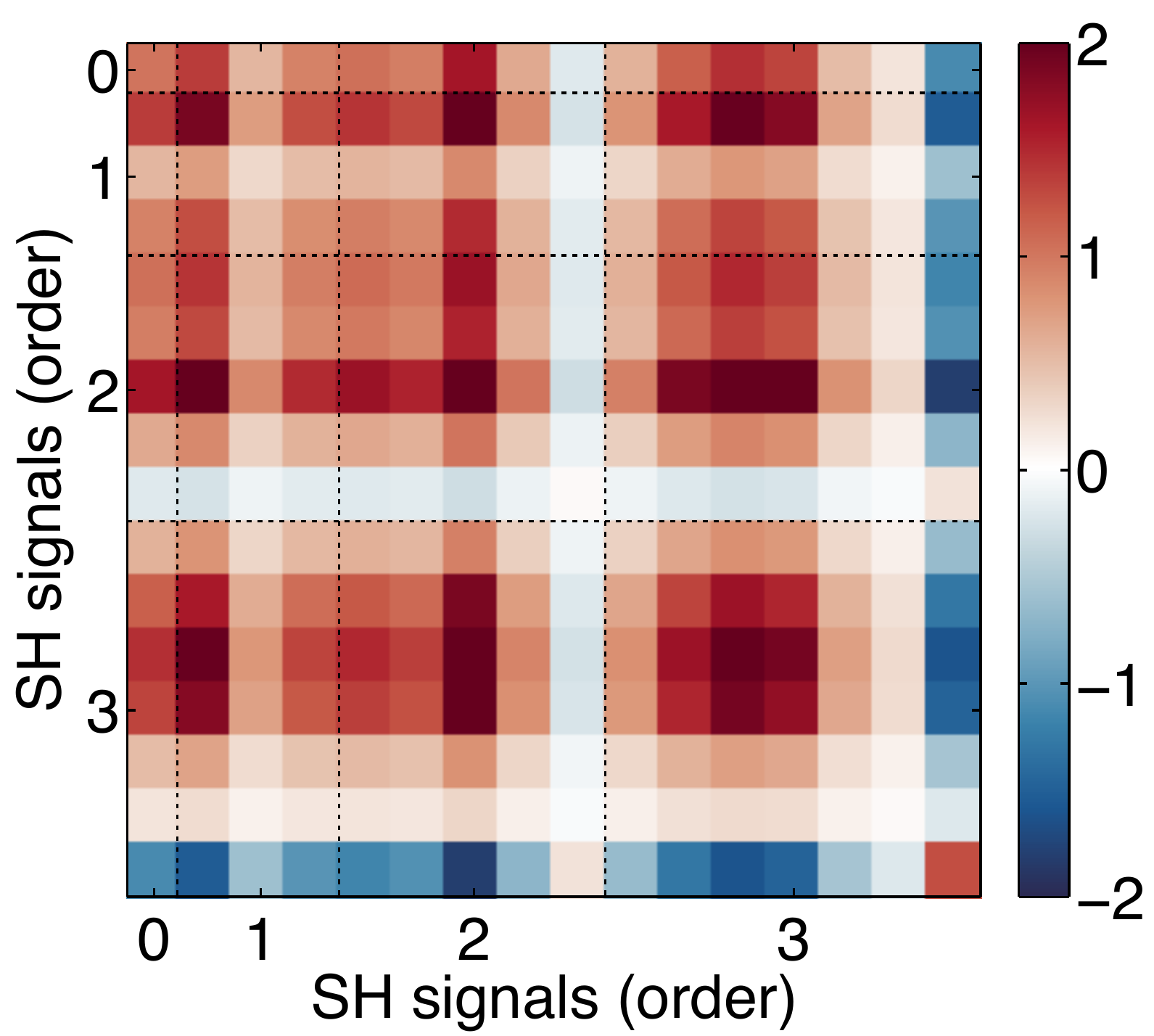}
& \includegraphics[width=.39\columnwidth]{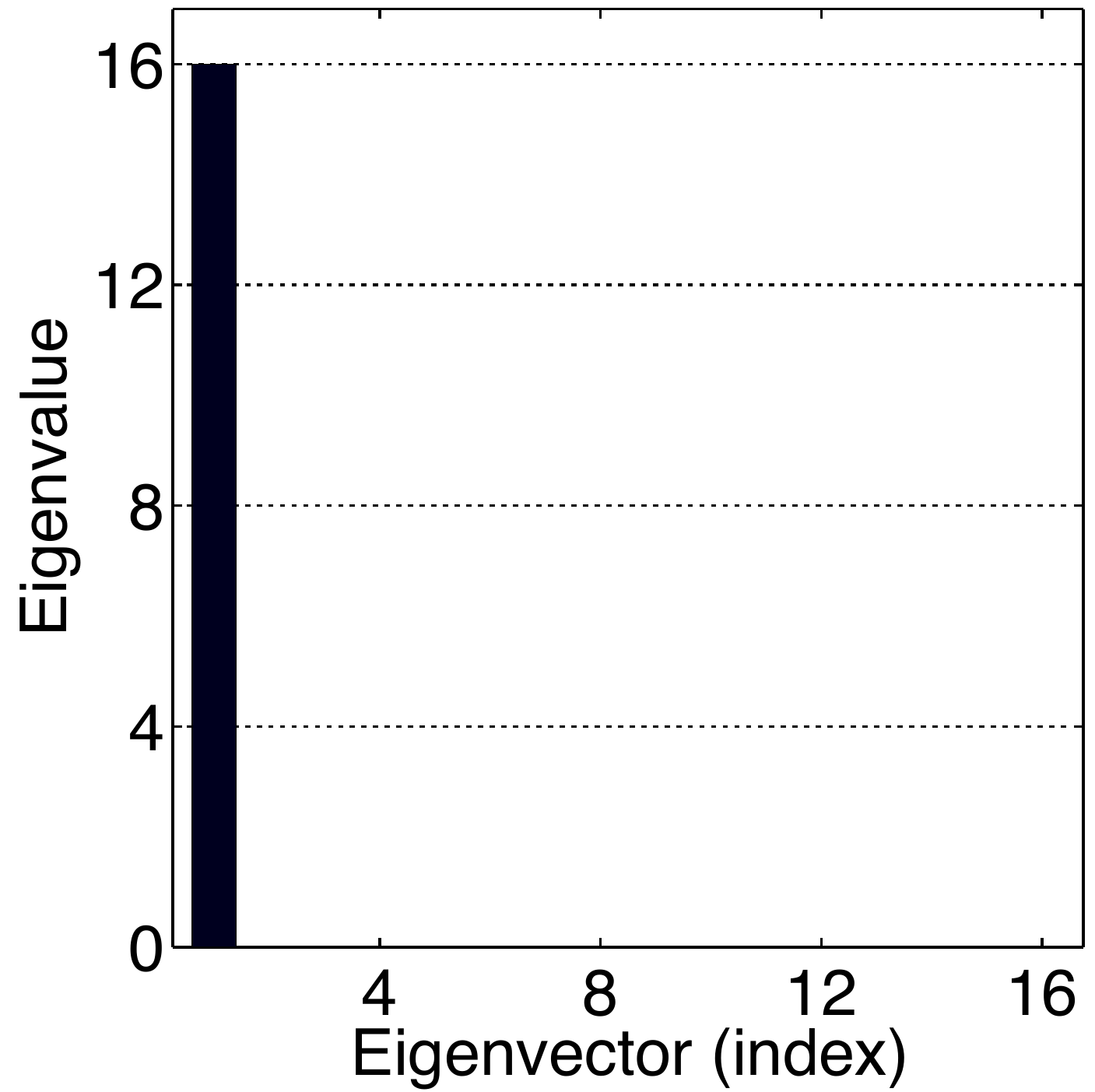} \\
\begin{sideways}\ \ \ \ \ \ \ \ \textsf{\small{$Q = 7$, $\beta=0.3$}}\end{sideways}
& \includegraphics[trim = 0mm .7mm 0mm 0mm, clip=true,width=.45\columnwidth]{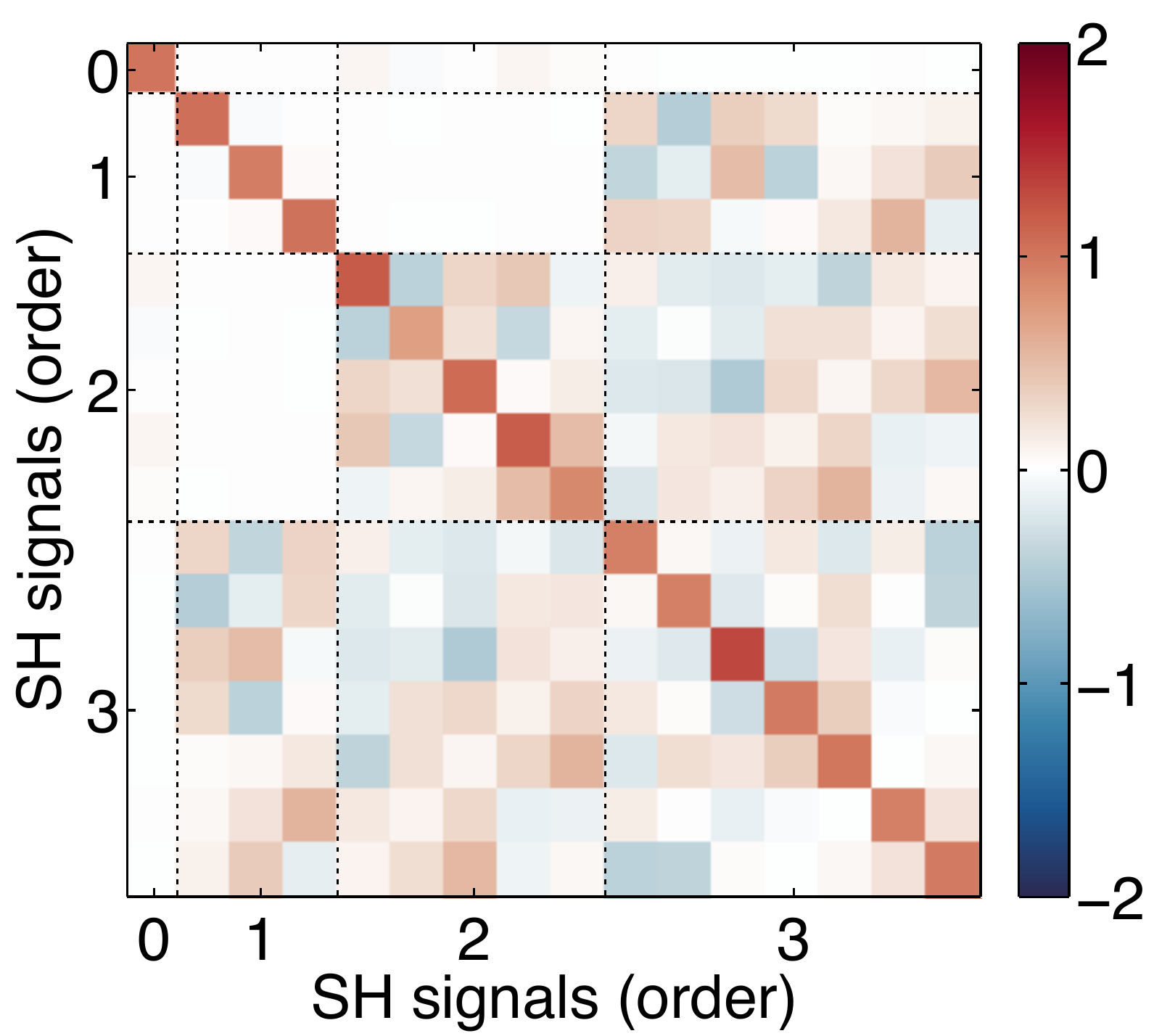}
& \includegraphics[width=.39\columnwidth]{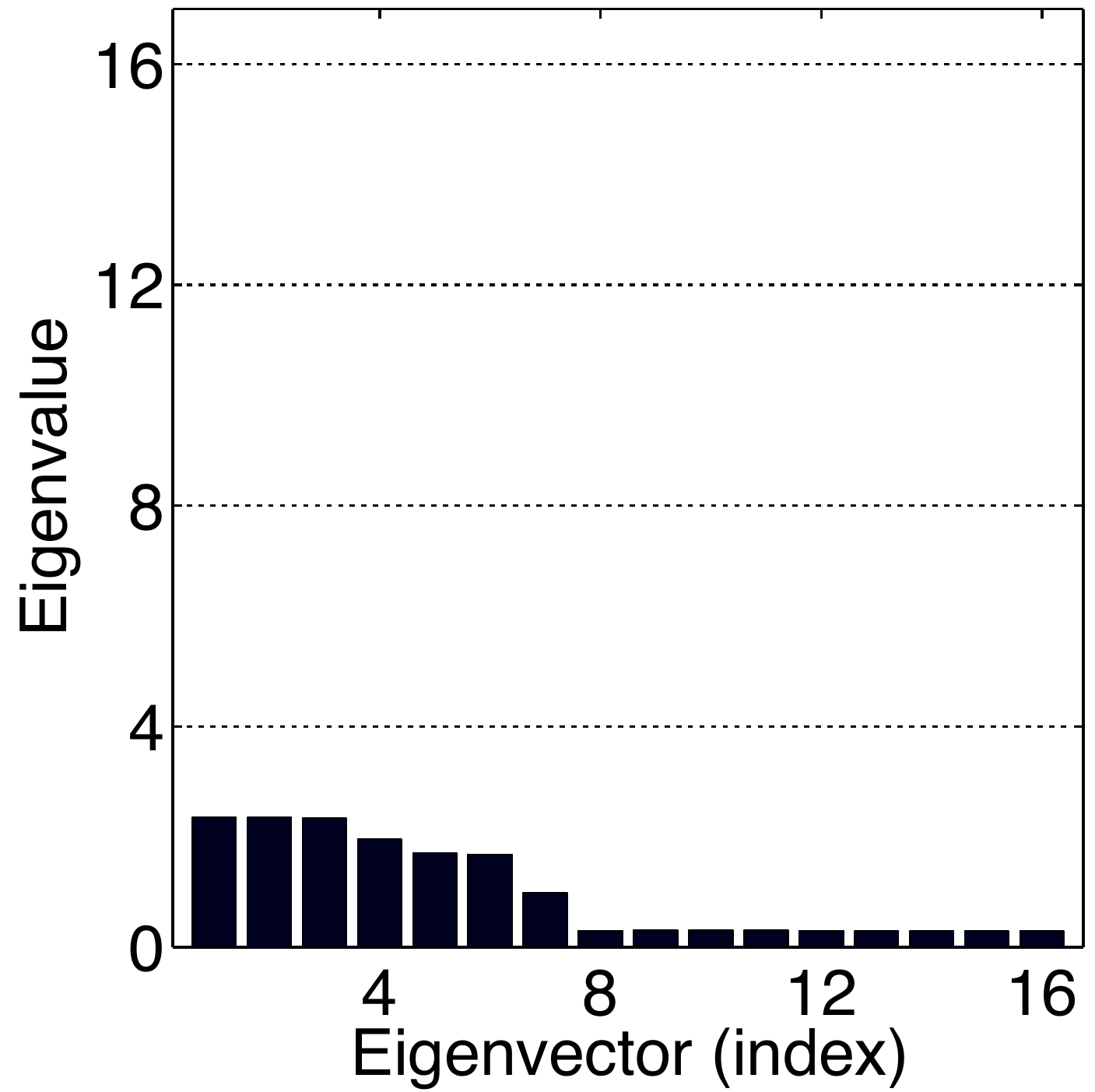} \\
\begin{sideways}\ \ \ \ \ \ \ \ \textsf{\small{$Q\rightarrow\infty$ / $\beta=1$}}\end{sideways}
& \includegraphics[trim = 0mm .7mm 0mm 0mm, clip=true,width=.45\columnwidth]{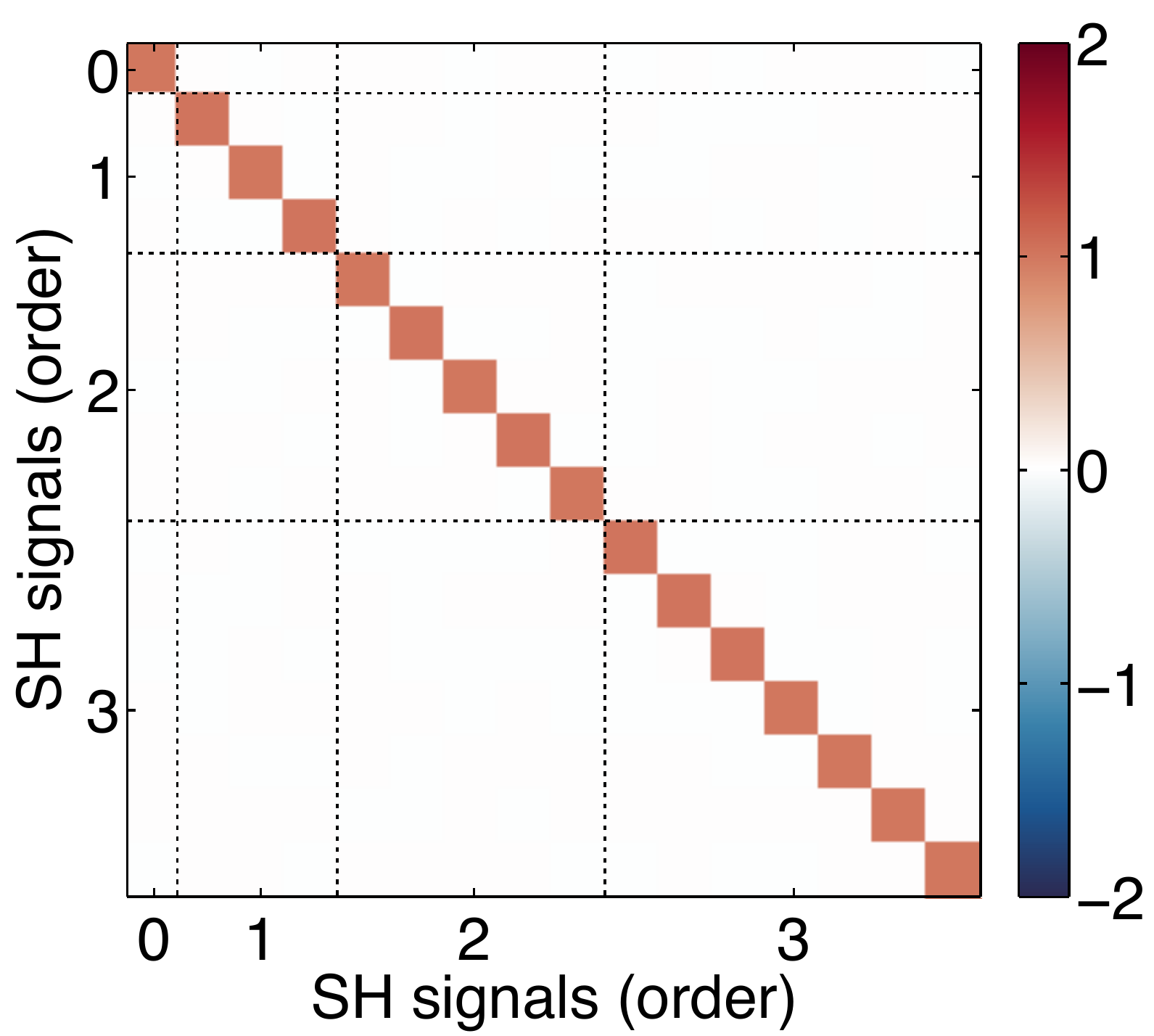}
& \includegraphics[width=.39\columnwidth]{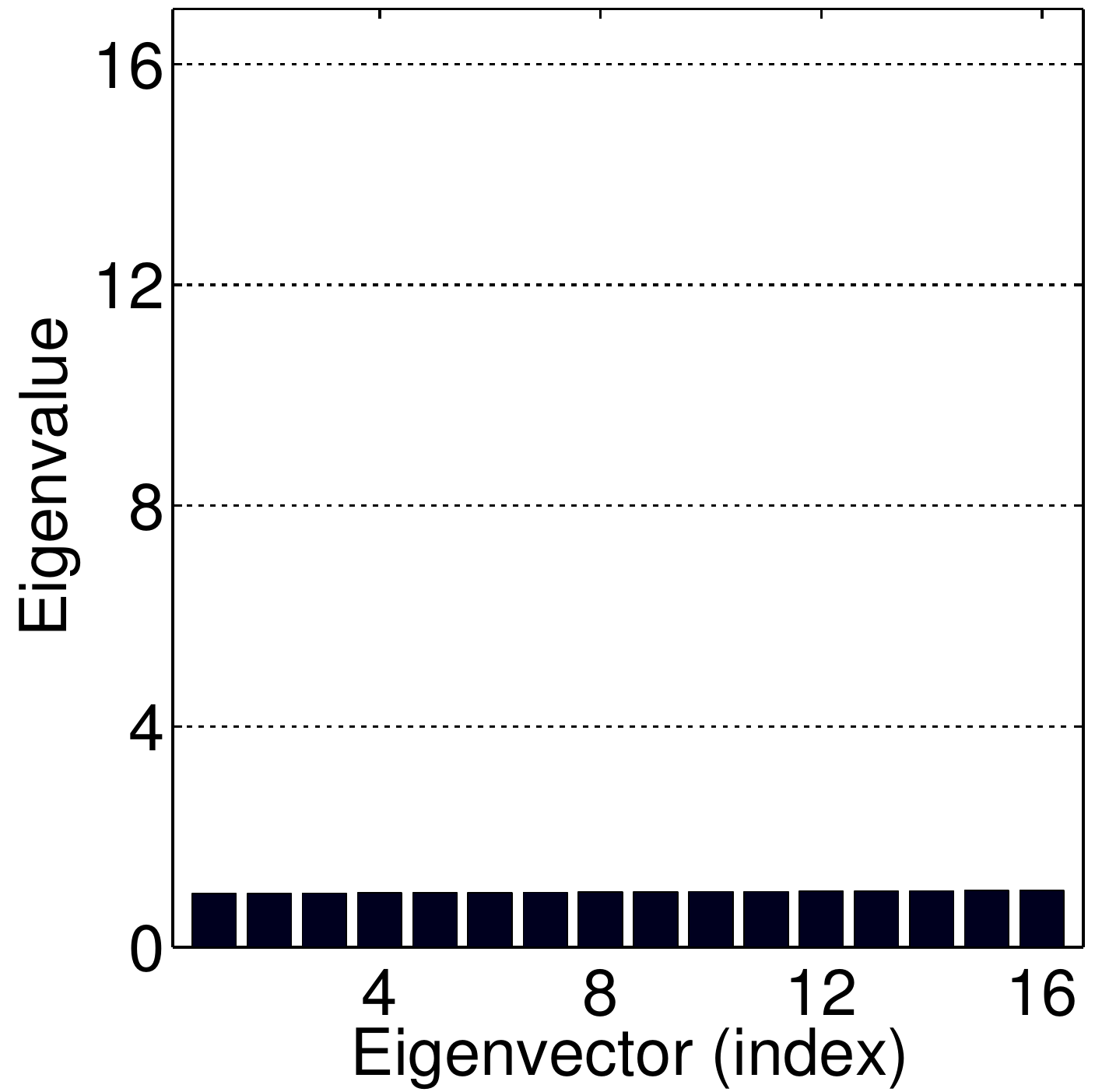} \\
\end{tabular}
\caption{This figure shows the SH signal covariance matrices (left) and the value of their respective eigenvalues (right) in three different situations: a purely directional sound field (top), a sound field consisting of several plane waves and a diffuse noise background (middle) and a perfectly diffuse sound field (bottom).}
\label{fig:CovMat}
\end{figure}

To provide some concreteness to these ideas, the \mbox{order-3} SH signal covariance matrix and its eigenvalue spectrum are illustrated in Figure~\ref{fig:CovMat} for three different situations: 1) a maximally non-diffuse sound field (resulting from one plane-wave); 2) a sound field consisting of several plane waves and a diffuse noise background; and 3) a perfectly diffuse sound field. In the case where the sound field consists of only one plane wave, the covariance matrix is rank-one and thus only the first eigenvalue is non-zero. On the contrary, when the sound field is perfectly diffuse, all of the eigenvalues are equal. Lastly, in the presence of a sound field consisting of several plane waves and a diffuse noise background, a more complex eigenvalue spectrum is observed, with the nine smallest eigenvalues corresponding to the purely diffuse component of the sound field and the seven largest ones corresponding to both the ``plane-wave'' and diffuse components. Clearly, the spectrum of the SH signal covariance matrix provides valuable information in characterizing the diffuseness of the sound field.

\section{Existing SH Diffuseness Measures}
\label{sec:otherMethods}
Before presenting our new SH diffuseness measure based on the spectrum of the SH covariance matrix, we briefly review two existing algorithms for estimating the diffuseness of sound fields in the SH domain and discuss their characteristics. 

\subsection{DirAC Diffuseness Measure for Order-1 SH Signals}
As briefly mentioned in the introduction, a possible way to measure the directionality or the diffuseness of the sound field is to compare the \emph{active} or propagating sound intensity with the energy density~\cite{Jacobsen1989, Schiffrer1994}. The calculation of the active sound intensity only requires the knowledge of the acoustic pressure and its derivatives along the $x$, $y$ and $z$ directions, which form the order-0 and order-1 SH signals, more commonly known as the B-format signals. This led Farina~\cite{Farina1999} to propose the calculation of what he referred to as the \emph{propagation index vector}, which was essentially the ratio of the active sound intensity vector to the total acoustic energy density, to improve the reconstruction of sound fields recorded in B-format. This idea was later rediscovered by Merimaa and Pulkki~\cite{Merimaa2005} for the analysis of B-format room impulse responses. Shortly after, Hurtado-Huyssen~\cite{Hurtado2005} and then Pulkki~\cite{Pulkki2007} proposed to use this diffuseness index for the reproduction of sound fields. As the active intensity ratio measure was made popular by Pulkki's directional audio coding (DirAC) method, we refer to it as simply the DirAC diffuseness measure.

The DirAC diffuseness, $\psi$, of the sound field is given by the relative flow of propagating acoustic energy traveling though the origin over time~\cite{Pulkki2007}:
\begin{equation}
\psi = 1-\frac{\left\| \bar{\mathbf{i}} \right\|}{c\,\bar{\varepsilon} } \ \ \text{,}
\label{eq:DirACDiffuseness}
\end{equation}
where $\bar{\mathbf{i}}$ denotes the time-averaged active sound intensity vector, $c$ denotes the speed of sound and $\bar{\varepsilon}$ denotes the time-averaged total energy of the sound field. In the presence of a single plane wave, $\left\|\bar{\mathbf{i}} \right\|=c\bar{\varepsilon} $ and $\psi$ is equal to zero. The time-averaged intensity and energy can be calculated from the SH signals using the following formulas:
\begin{align}
\bar{\varepsilon} &=\sum_{l=0}^{1}\sum_{m=-l}^{l}\mathrm{E}\left\{ b_{l,m}^2(t)\right\} \nonumber \\
\bar{\mathbf{i}} &=\frac{4c}{\sqrt{3}}
\left[ \, \mathrm{E}\left\{ b_{0,0}(t)\,b_{1,1}(t)\right\}\mathbf{v}_x \ \hdots \right. \nonumber \\
& \left. +\,\mathrm{E}\left\{ b_{0,0}(t)\,b_{1,-1}(t)\right\}\mathbf{v}_y +\mathrm{E}\left\{ b_{0,0}(t)\,b_{1,0}(t)\right\}\mathbf{v}_z \, \right]
\label{eq:DiracEnergyIntensity}
\end{align}
where $\mathbf{v}_x$, $\mathbf{v}_y$ and $\mathbf{v}_z$ represent the unit vectors in the directions of the corresponding cartesian coordinate axes. Note that these formulas differ from those given in~\cite{Pulkki2007}. This is because the expressions in Equation~\eqref{eq:DiracEnergyIntensity} are given for fully-normalized spherical harmonics whereas the formulas in~\cite{Pulkki2007} are given for 2D-semi-normalized spherical harmonics. Please refer to~\cite{Daniel2003} for more details on the normalization of SH signals.

Two main remarks can be made about the DirAC diffuseness measure. First, it uses the SH signals up to order~1 only, therefore it does not take full advantage of the spatial resolution offered by an SMA with a large number of sensors. We saw in Section~\ref{subsec:shCov} that a sound field that appeared very diffuse at order~1 did not necessarily seem so diffuse when considering order-4 SH signals (see Figure~\ref{fig:DirectDiffuseTransition} for $Q=10$, for instance).

%The DirAC order-1 diffuseness measure has two major drawbacks. First, it uses the SH signals up to order~1 only. Therefore, it provides a very \emph{local} diffuseness estimate and does not take advantage of the additional spatial information acquired by an SMA with a large number of sensors. For instance, SH signals up to order~3 provide a description of the sound field over an area that is much larger than that described by only order-1 signals. A sound field that seems very diffuse when considering order-1 SH signals may not seem so diffuse when considering higher-order SH signals.

Second, in the presence of two sources located in opposite directions and having equal powers, the DirAC algorithm yields a diffuseness estimate corresponding to a perfectly diffuse sound field regardless of the source signal correlation. To see this, consider the SH representation of two sources located in opposite directions:
\begin{equation}
\begin{cases}
b_{0,0}(t) = s_1(t) + s_2(t) \\
b_{1,m}(t)=Y_1^m(\Omega_1)\left ( s_1(t) - s_2(t)\right ) \ \text{,}
\end{cases}
\end{equation}
where $\Omega_1$ denotes the direction of the first source and $s_i(t)$ denotes the $i$-th source signal. 
The average intensity vector, $\bar{\mathbf{i}}$, is then given by:
\begin{align}
\bar{\mathbf{i}} & =\frac{4c}{\sqrt{3}T} \left( \mathrm{E}\left\{ s_1^2(t)\right\} -\mathrm{E}\left\{ s_2^2(t) \right\}\right) \dots \nonumber \\ 
&\ \ \ \times \left [Y_1^{1}(\Omega_1)\,\mathbf{v}_x +Y_1^{-1}(\Omega_1)\,\mathbf{v}_y+Y_1^{0}(\Omega_1)\,\mathbf{v}_z\right ]
\end{align}
If we now assume that the power of the source signals are equal, we have $\mathrm{E}\left\{ s_1^2(t)\right\} =\mathrm{E}\left\{ s_2^2(t) \right\}$. The average intensity vector is then equal to $\mathbf{0}$ and $\psi=1$. In the following we show experimentally that the DirAC diffuseness is also close to $1$ whenever $Q$ sources with equal power are evenly distributed over the sphere and are either mutually uncorrelated or have identical signals. In other words, the DirAC diffuseness estimator has difficulty characterizing sound field diffuseness that arises from the presence of multiple sources located in opposite directions.

Note that, in a recent work~\cite{Pulkki2013}, Pulkki described a generalization of the DirAC method for SH signals of arbitrarily high order. In this work, the high-order SH signals are first converted to order-1 SH signals corresponding to different angular sectors. The diffuseness of the sound field is then calculated using Equation~\eqref{eq:DirACDiffuseness} for each angular sector. Two remarks can be made regarding this method. First, the concept of angle-dependent diffuseness is difficult to interpret, because diffuseness classically measures how evenly the acoustic energy is flowing from every direction in space. Second, similar to the order-1 DirAC diffuseness measure, the diffuseness measured for a particular angular sector can be equal to one in the presence of only two sources. This can happen for a sector centered around the $x$-axis, for instance, if the two sources have equal power and are located in opposite directions along the $y$-axis.

\subsection{Thiele-Gover Diffuseness Measure}
Another approach for measuring diffuseness is that of ``directional diffusion''~\cite{Kuttruff2009} which consists in analyzing the directional distribution of sound energy flow. The method is based on the notion that the more homogeneous this distribution, the more diffuse the sound field. The method was first proposed by Meyer and Thiele~\cite{Meyer1956}, who manually rotated a directional microphone in many directions to measure the corresponding amounts of acoustic energy. More recently, Gover~\cite{Gover2002,Gover2004} adapted the method to SMAs and proposed to measure the directional distribution of acoustic energy by steering directional beams in every direction of space. In this context the energy, $e(\Omega_n)$, incoming from direction $\Omega_n$ is estimated as: 
\begin{equation}
e(\Omega_n) =\mathrm{E} \left \{ \hat{s}_n^2(t) \right \}\ \text{,}
\label{eq:energyDistrib}
\end{equation}
where $\hat{s}_n(t)$ is the output of a spherical beamformer steered in the direction $\Omega_n$, \emph{i.e.}:
\begin{equation}
\hat{s}_n(t) =\frac{1}{(L+1)^2}\, \mathbf{y}(\Omega_n)^\mathsf{T}\, \mathbf{b}(t)\ \text{.}
\end{equation}
Note that this expression corresponds to the spherical beamformer with maximum directivity. This measurement is repeated for a large number, $N$, of directions evenly distributed over the sphere. The variation of energy across directions is then measured as the average deviation, $\mu$, of the energy from its mean: 
\begin{align}
& \mu = \frac{1}{\langle e \rangle}\sum_{n=1}^{N}\left|e(\Omega_n)-\langle e \rangle\right|\ \text{,} \nonumber \\
& \text{where } \langle e \rangle = \frac{1}{N}\sum_{n=1}^{N}e(\Omega_n)\ \text{.}
\end{align}
The Thiele-Gover diffuseness, $\phi$, of the sound field is then estimated from the mean energy deviation using the following formula: 
\begin{equation}
\phi = 1 - \frac{\mu}{\mu_0} \ \text{,}
\end{equation}
where $\mu_0$ is the value of $\mu$ in the most non-diffuse case, \emph{i.e.} in the presence of a single plane wave. If the energy is evenly distributed across directions, $\mu=0$ thus $\phi=1$ and the sound field is interpreted as perfectly diffuse. On the contrary, if only one plane wave is present, $\mu=\mu_0$ resulting in $\phi=0$ and the sound field is interpreted as perfectly directional.

The Thiele-Gover diffuseness measure presents very different characteristics to that of the DirAC diffuseness. First, the spatial information contained in the SH signals with order greater than one is used. Second, in the presence of two sources located in opposite directions, the energy distribution is not homogeneous and therefore the sound field is not misinterpreted as perfectly diffuse. Therefore, the Thiele-Gover diffuseness is able to characterize sound field diffuseness arising from the presence of multiple sources located in opposite directions.

A characteristic of this algorithm, however, is that is is based on the directional energy only and disregards the correlation between signals incoming from different directions. For example, an exceptional situation arises when the sound field consists of a large number of plane waves that are evenly distributed across space and carrying the same signal. In this case, the plane waves are perfectly in phase at the center of the SMA resulting in a sound pressure at the center that is much larger than the sound pressure at other locations. This sound field clearly does not fit the definition of a diffuse sound field, yet it is interpreted as perfectly diffuse using the Thiele-Gover diffuseness measure because it provides a perfectly even directional energy distribution. 

\section{The COMEDIE Diffuseness Estimator}
\label{sec:myMethod}
In this section we present a new SH diffuseness measure, the COMEDIE (covariance matrix eigenvalue diffuseness estimation) diffuseness estimator. 

\subsection{Algorithm}
In section~\ref{subsec:CovMat} we have shown that the spectrum of the SH signal covariance matrix is strongly related to the diffuseness of the sound field. Specifically, we have shown that the eigenvalues of the covariance matrix are most similar in the presence of a diffuse sound field, and most dissimilar in the presence of a single plane wave. Therefore, we propose to estimate the diffuseness of the sound field based on the homogeneity of the SH signal covariance matrix spectrum.

Similar to the Thiele-Gover algorithm, we estimate the COMEDIE diffuseness, $d$, of the sound field as:
\begin{equation}
d = 1 - \frac{\gamma}{\gamma_0} \ \text{,}
\end{equation}
where:
\begin{itemize}
\item $\gamma$ is the deviation of the eigenvalues of the SH signal covariance matrix from their mean, \emph{i.e.}:
\begin{align}
& \gamma = \frac{1}{\langle v \rangle}\sum_{i=1}^{(L+1)^2}\left|v_{i}-\langle v \rangle\right|\ \text{,} \nonumber \\
& \text{where }\langle v \rangle= \frac{1}{(L+1)^2}\sum_{i=1}^{(L+1)^2}v_{i}\ \text{,}
\label{eq:eigVar}
\end{align}
\item $\gamma_0$ is the value of $\gamma$ in the most non-diffuse case, that is in the presence of a single plane wave with $\beta=0$ and is given by (derived below, see Eq.~\ref{eq:nondiffuse}):
\begin{equation}
\gamma_0 = 2\left[(L+1)^2-1\right]\ \text{.}
\end{equation}
\end{itemize}

We now prove that in the presence of a sound field consisting of a single plane wave ($Q=1$) and a diffuse noise background, the COMEDIE diffuseness, $d$, is equal to the relative noise level, $\beta$. This is a desirable characteristic, because $\beta$ corresponds to the definition commonly found in the literature for diffuseness when only one source is present. Furthermore we show in the next section that, in the presence of a single source, the DirAC and the Thiele-Gover diffuseness estimates are equal to the COMEDIE estimate of diffuseness. Without loss of generality we assume that the total power of the SH signals in the absence of noise ($\beta=0$) is equal to $(L+1)^2$. The relative noise level in the SH signals is then given by:
\begin{equation}
\beta = \frac{(L+1)^2\nu}{(L+1)^2+(L+1)^2\nu} = \frac{\nu}{1+\nu} \ \text{.}
\end{equation}
Therefore, we have:
\begin{equation}
\nu=\frac{\beta}{1-\beta}\ \text{.}
\end{equation}
The directional component of the sound field originates from a single plane wave, therefore the entire power of this component concentrates in the first eigenvalue of the SH signal covariance matrix. In other words, using the notations of Section~\ref{sec:Model} we have:
 \begin{equation}
\begin{cases}
w_1 = (L+1)^2 \\
w_i = 0\ \ \ \  \forall \ i > 1 \ \text{.}
\end{cases}
\end{equation}
On the other hand, the power of the diffuse noise component distributes evenly to each eigenvalue. Thus, according to Equation~\eqref{eq:CovSpec}, the eigenvalues of the total SH signal covariance matrix are given by:
\begin{equation}
\begin{cases}
v_{1} = \frac{\beta}{1-\beta}+(L+1)^2 \\
v_{i} = \frac{\beta}{1-\beta}\ \ \ \ \forall \ i > 1\ \text{.}
 \end{cases}
\end{equation}
And we have:
\begin{align}
\langle v \rangle&= \frac{1}{(L+1)^2}\left((L+1)^2\frac{\beta}{1-\beta}+(L+1)^2\right) \nonumber \\ 
&= \frac{\beta}{(1-\beta)}+\frac{1-\beta}{(1-\beta)} \nonumber \\
&= \frac{1}{(1-\beta)}\ \text{,} \nonumber \\
\gamma & = (1-\beta)\left[ \left((L+1)^2-1\right)\left |\frac{\beta}{1-\beta}-\frac{1}{1-\beta} \right |\dots \right. \nonumber \\
& \ \ \ \ \ \ \ \ \ \ \ \ \ \ \ \ \ \left. +\left| (L+1)^2+\frac{\beta}{1-\beta}-\frac{1}{1-\beta}\right|\ \right] \nonumber \\
& = (1-\beta)\left[ (L+1)^2-1+(L+1)^2-1\right] \nonumber \\
& =2\,(1-\beta)\left[ (L+1)^2-1\right]\ \text{.}
\end{align}
Setting $\beta$ to 0 we obtain $\gamma_0$, the value of $\gamma$ in the most non-diffuse case:
\begin{equation}
\gamma_0 = 2\left[(L+1)^2-1\right]\ \text{.}
\label{eq:nondiffuse}
\end{equation}
Therefore, $d$ is given by:
\begin{align}
d &= 1-\frac{2\left[(L+1)^2-1\right](1-\beta)}{2\left[(L+1)^2-1\right]} \nonumber \\
& =1-(1-\beta) \nonumber \\ 
& =\beta\ \text{.}
\end{align}

Because the COMEDIE algorithm is based on covariance, it does not interpret a sound field consisting of many perfectly correlated plane waves distributed over the sphere as perfectly diffuse, contrary to the Thiele-Gover method. In this situation, the rank of the covariance matrix is one and thus $d$ is equal to $0$. More generally, it can be seen that the COMEDIE diffuseness estimate takes into consideration both the spatial distribution and correlation between signals.

Note that in the derivations above we implicitly assumed that the SH signals are free of measurement noise. In practice, SH signals recorded with an SMA are always noisy, mostly due to measurement noise at low frequency and spatial aliasing at high frequency~\cite{Rafaely2005}. The diffuseness of the sound field can be accurately measured only if the SH signals are relatively clean, which can be achieved over a wide frequency range using a properly designed SMA~\cite{Jin2013}. As measurement noise is generally spatially white and uncorrelated across sensors, it increases the estimated diffuseness. This remark applies to all of the methods described in this paper, \emph{i.e.} for the COMEDIE algorithm as well as the other methods.

\subsection{Numerical Simulations}
\label{subsec:Simul}
In order to illustrate the behavior of the COMEDIE diffuseness estimator, we present the results of some simple numerical simulations. In the first simulation we model partially diffuse sound fields as described in Section~\ref{sec:Model}. The SH signals are simulated as the sum of a perfectly diffuse sound field component and a component consisting of $Q$ plane-waves. In every case, the length of the SH signals is 1024 samples and they are calculated using Equation~\eqref{eq:Model}. The noise signals and plane-wave signals are mutually uncorrelated Gaussian white noise signals with equal power and the plane-wave directions form a $Q$-point sphere packing~\cite{SloanePackings}. We estimate the diffuseness of the sound field as a function of $Q$ for three different diffuseness measures: the order-1 DirAC diffuseness measure, the Thiele-Gover directional diffuseness and the COMEDIE diffuseness estimator.

\begin{figure}[t]
\centerline{
\includegraphics[width=.48\columnwidth]{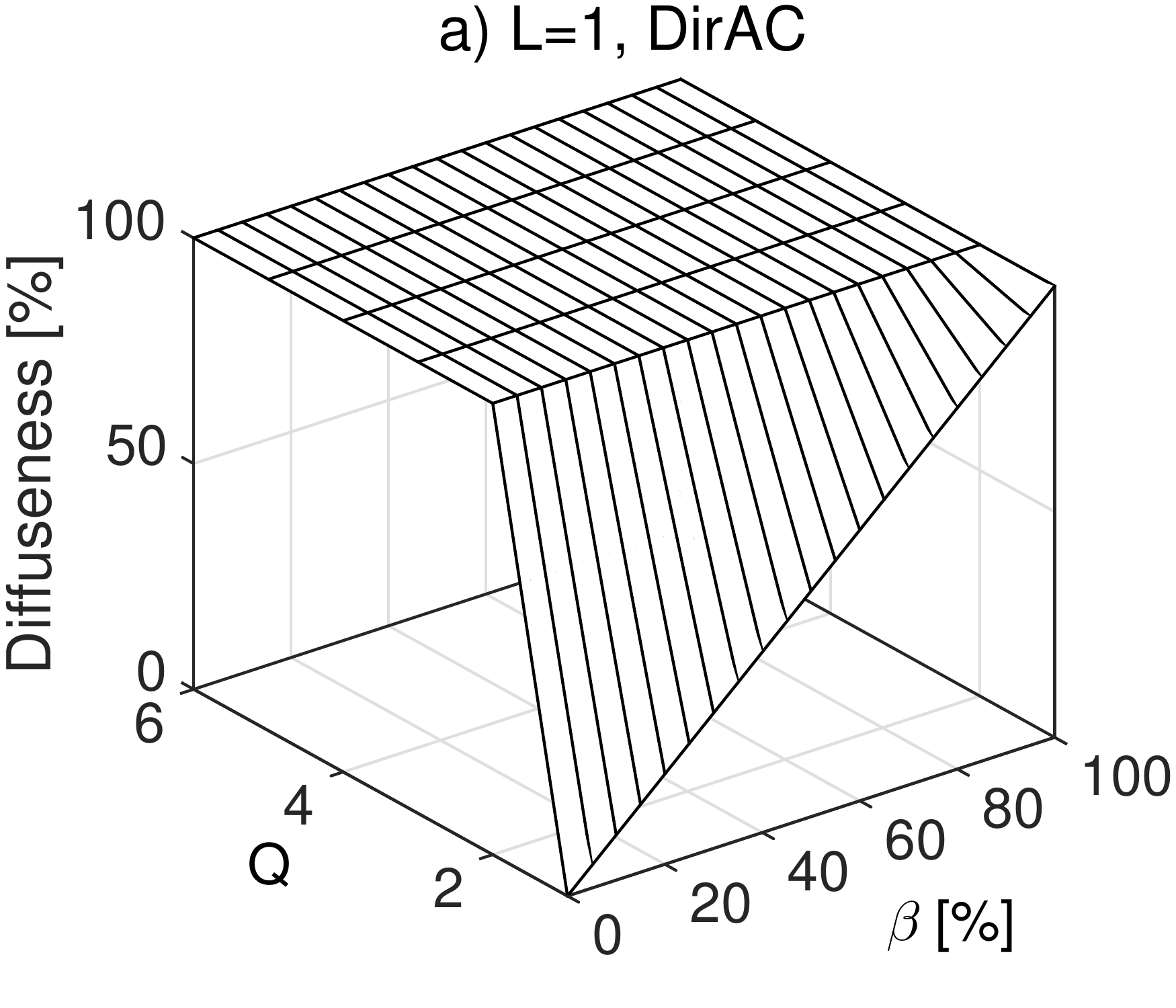}
}
\centerline{
\includegraphics[width=.48\columnwidth]{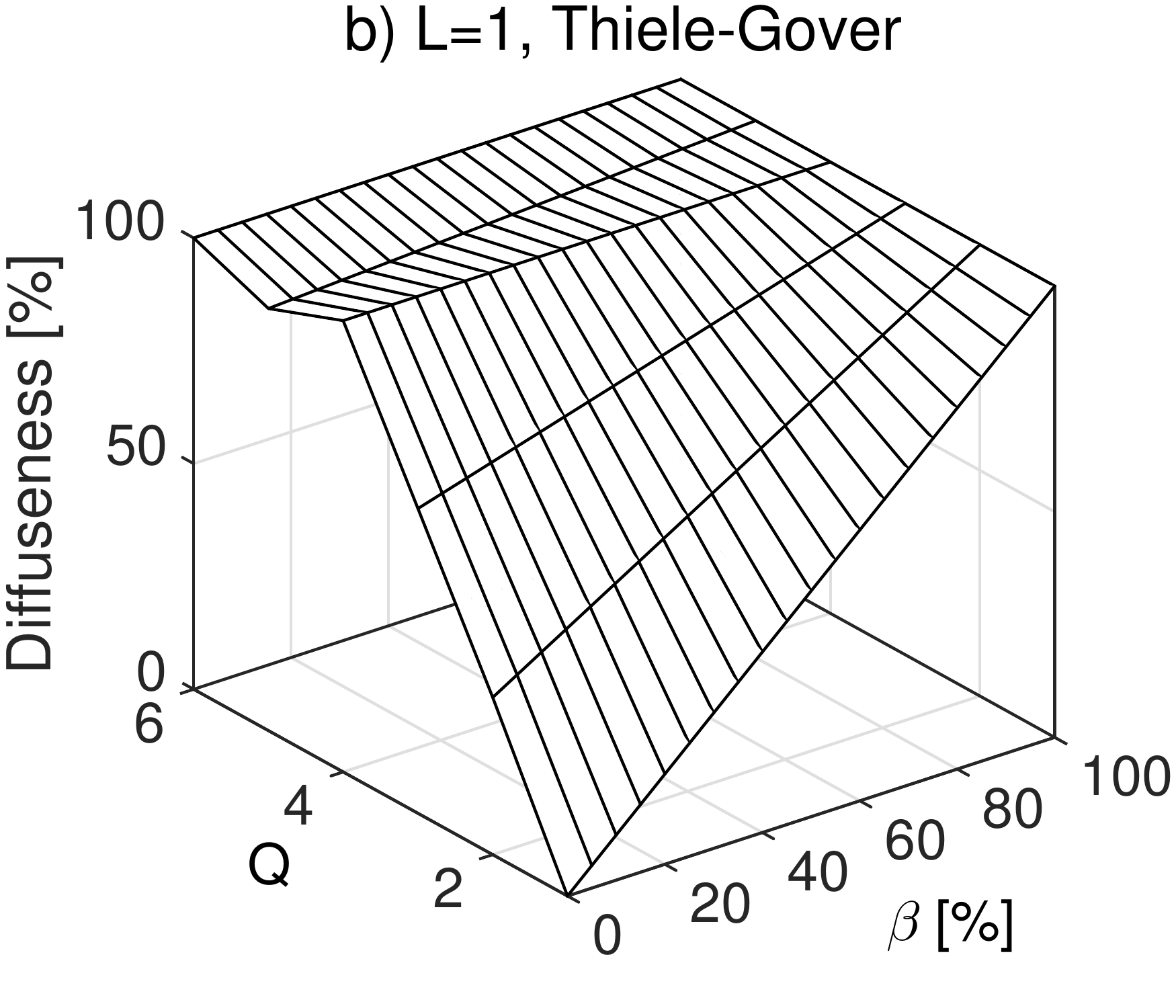}
\includegraphics[width=.48\columnwidth]{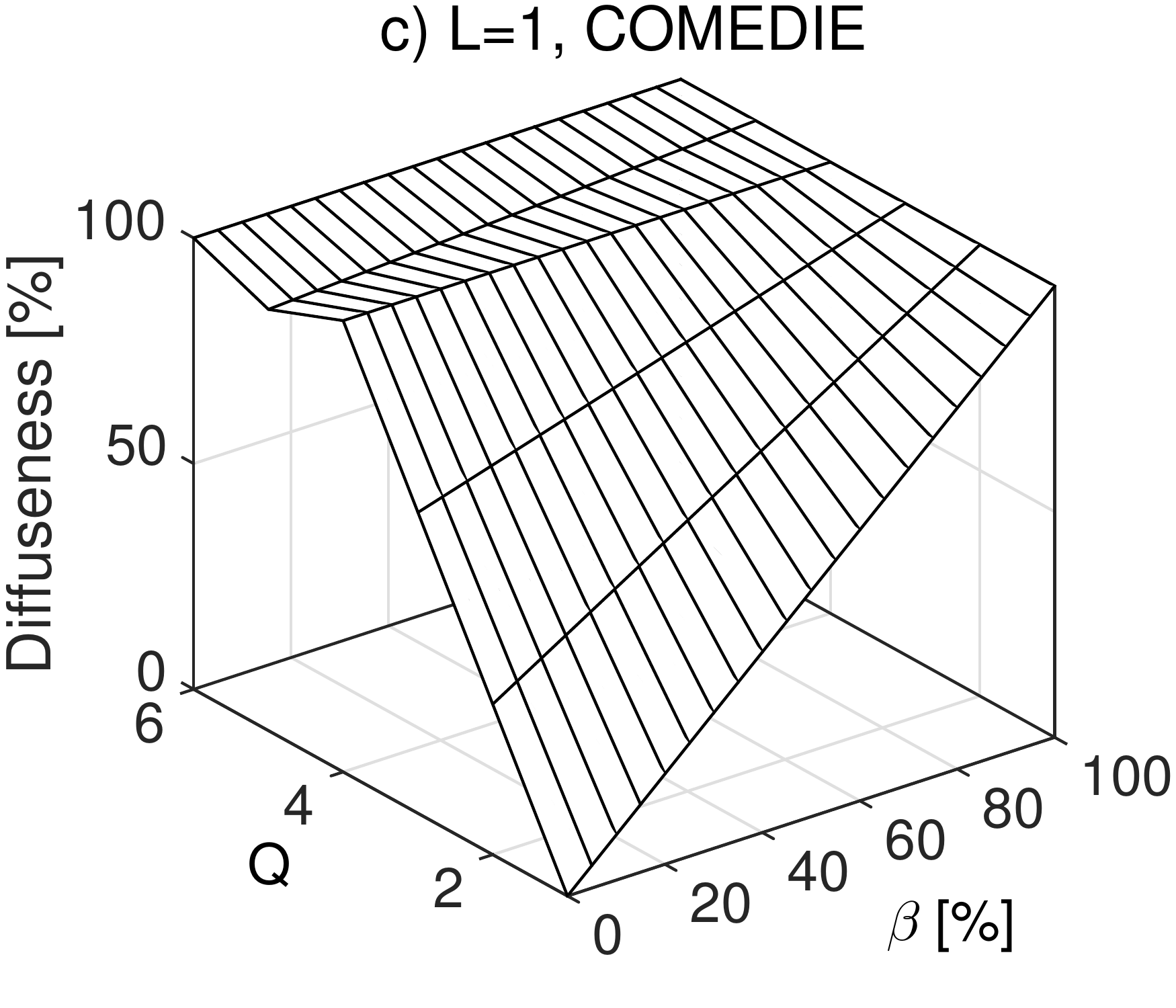}
} 
\centerline{
\includegraphics[width=.48\columnwidth]{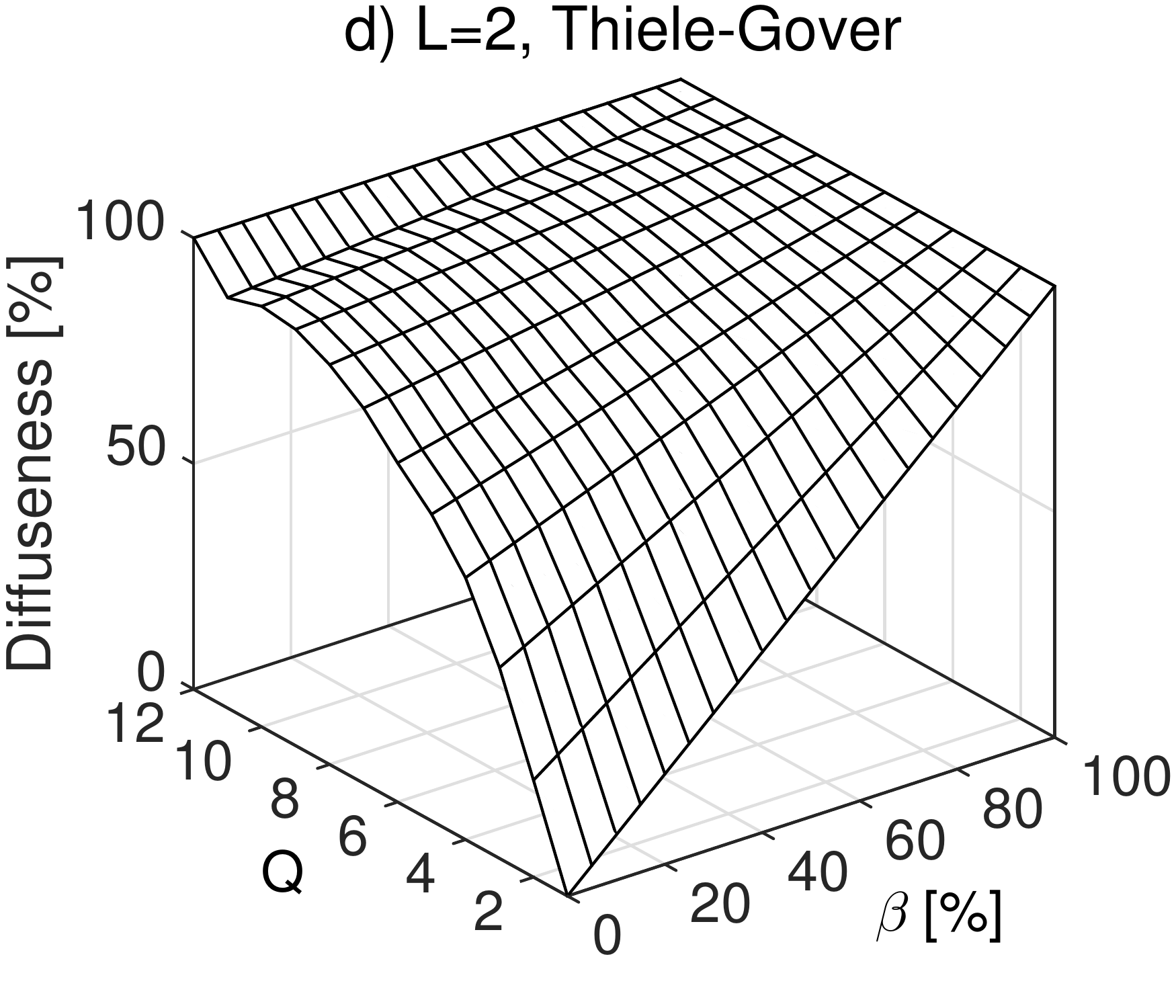}
\includegraphics[width=.48\columnwidth]{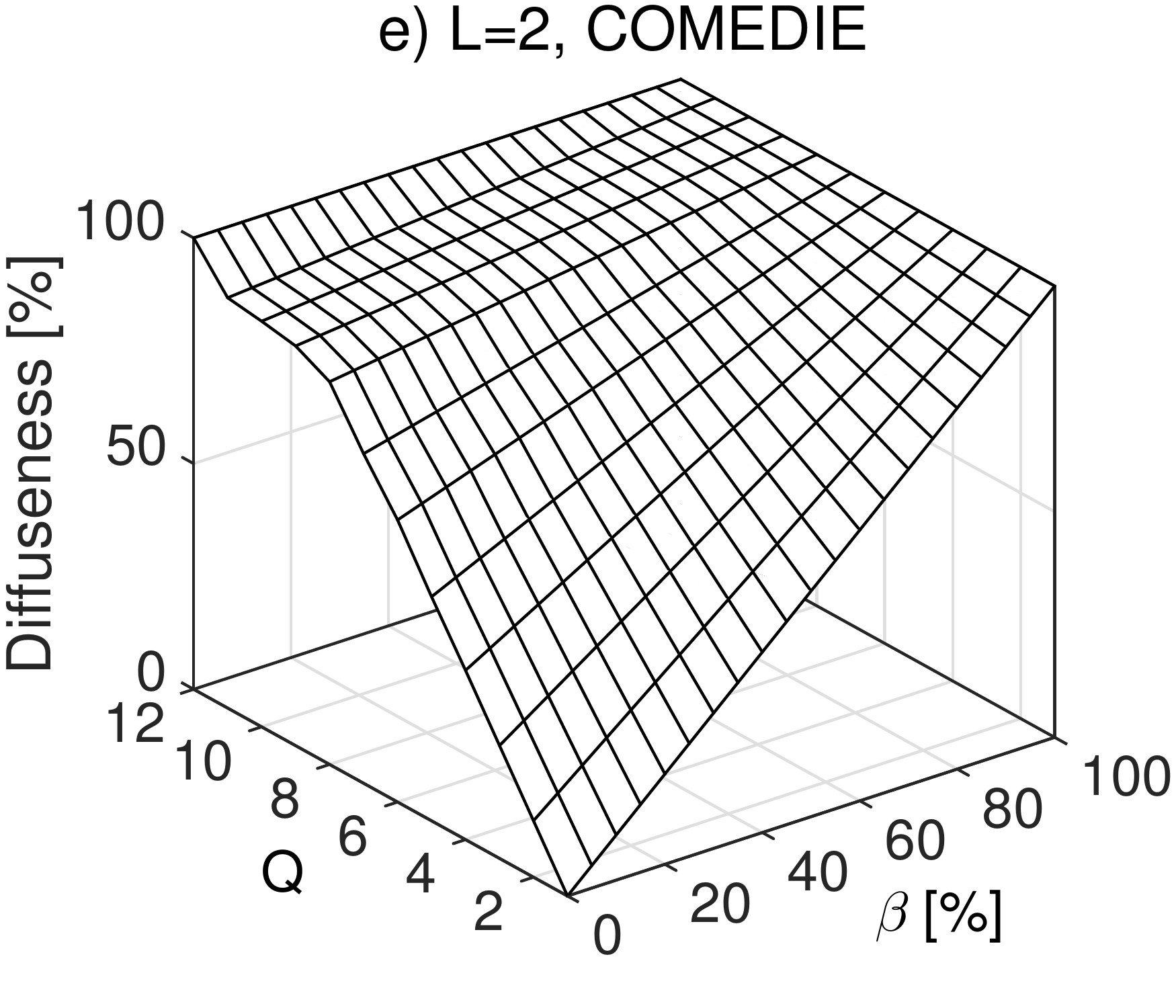}
} 
\centerline{
\includegraphics[width=.48\columnwidth]{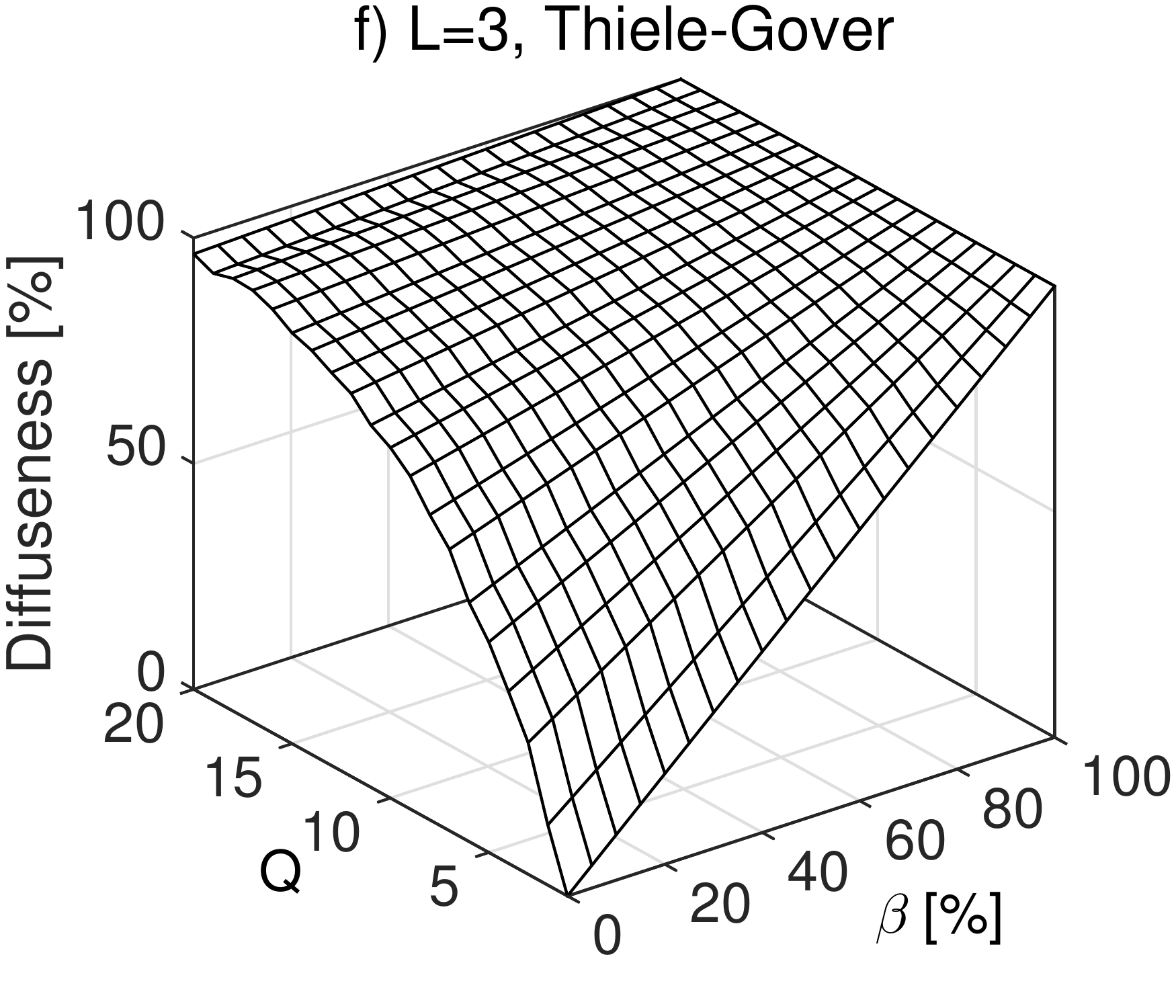}
\includegraphics[width=.48\columnwidth]{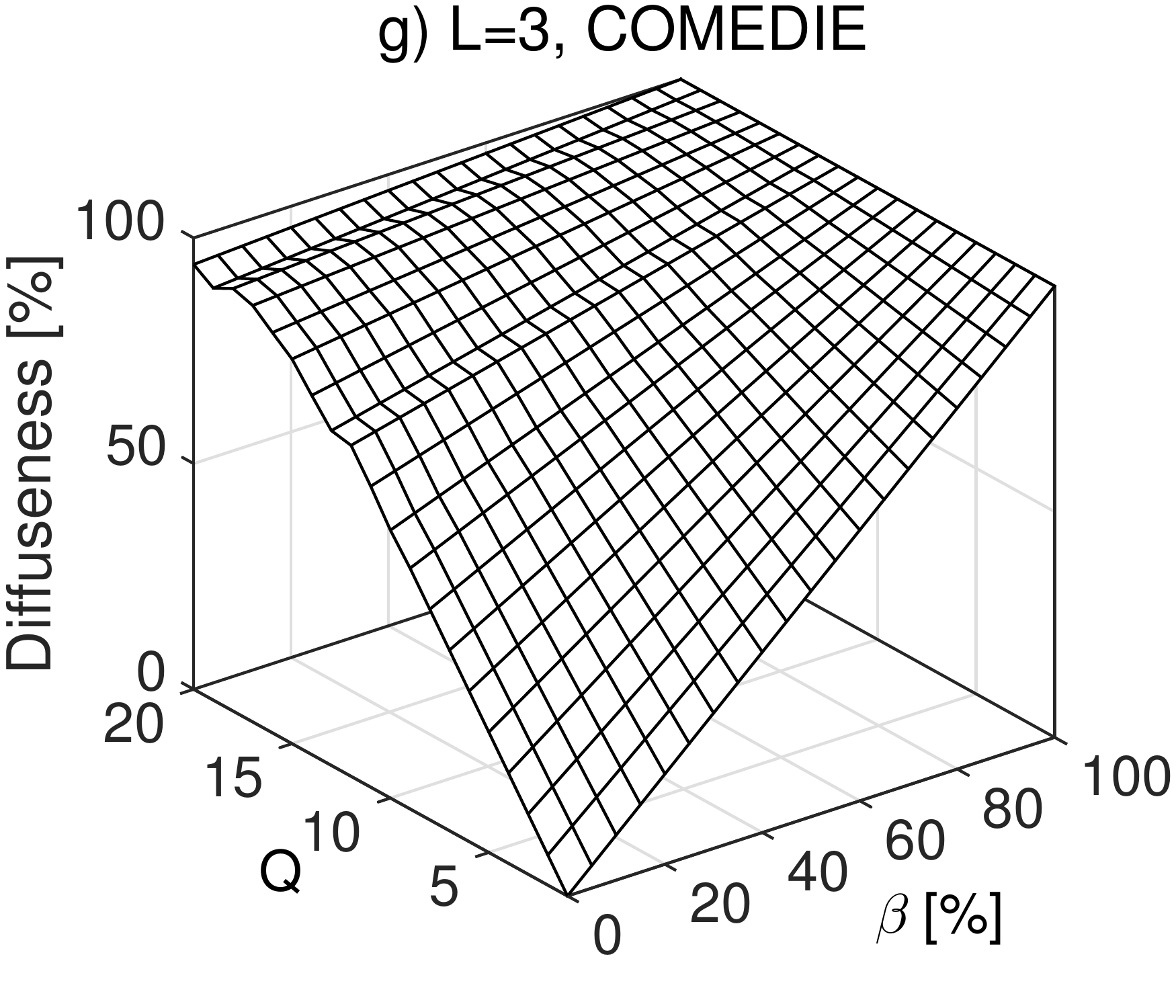}
} 
%\vspace{2mm}
%\centerline{\includegraphics[width=.48\columnwidth]{./figures/colorbarHoriz.pdf}}
\caption{This figure shows the estimated diffuseness in the presence of uncorrelated sound sources and a perfectly diffuse component. The plots represent the estimated diffuseness as a function of the number of sound sources and diffuse sound field relative energy, using: a) The DirAC method with order-1 SH signals; b-d-f) The Thiele-Gover method applied to SH signals with orders 1, 2 and 3; c-e-g) the COMEDIE method applied to SH signals with orders 1, 2 and 3.}
\label{fig:CompNumSou}
\end{figure}

%\begin{figure}[t]
%\centerline{
%\includegraphics[width=.48\columnwidth]{./figures/SouSpn_Plk01.pdf}
%}
%\centerline{
%\includegraphics[width=.48\columnwidth]{./figures/SouSpn_Gov01.pdf}
%\includegraphics[width=.48\columnwidth]{./figures/SouSpn_Hoa01.pdf}
%} 
%\centerline{
%\includegraphics[width=.48\columnwidth]{./figures/SouSpn_Gov02.pdf}
%\includegraphics[width=.48\columnwidth]{./figures/SouSpn_Hoa02.pdf}
%} 
%\centerline{
%\includegraphics[width=.48\columnwidth]{./figures/SouSpn_Gov03.pdf}
%\includegraphics[width=.48\columnwidth]{./figures/SouSpn_Hoa03.pdf}
%} \vspace{2mm}
%\centerline{
%\includegraphics[width=.48\columnwidth]{./figures/colorbarHoriz.pdf}
%}
%\caption{This figure shows the estimated diffuseness as a function of the angular span between sources and diffuse sound field relative energy, using: a) The DirAC method with order-1 SH signals; b-d-f) The Thiele-Gover method applied to SH signals with orders 1, 2 and 3; c-e-g) the COMEDIE method applied to SH signals with orders 1, 2 and 3.}
%\label{fig:CompAngSpan}
%\end{figure}

\begin{figure}[t]
\centerline{
\includegraphics[width=.48\columnwidth]{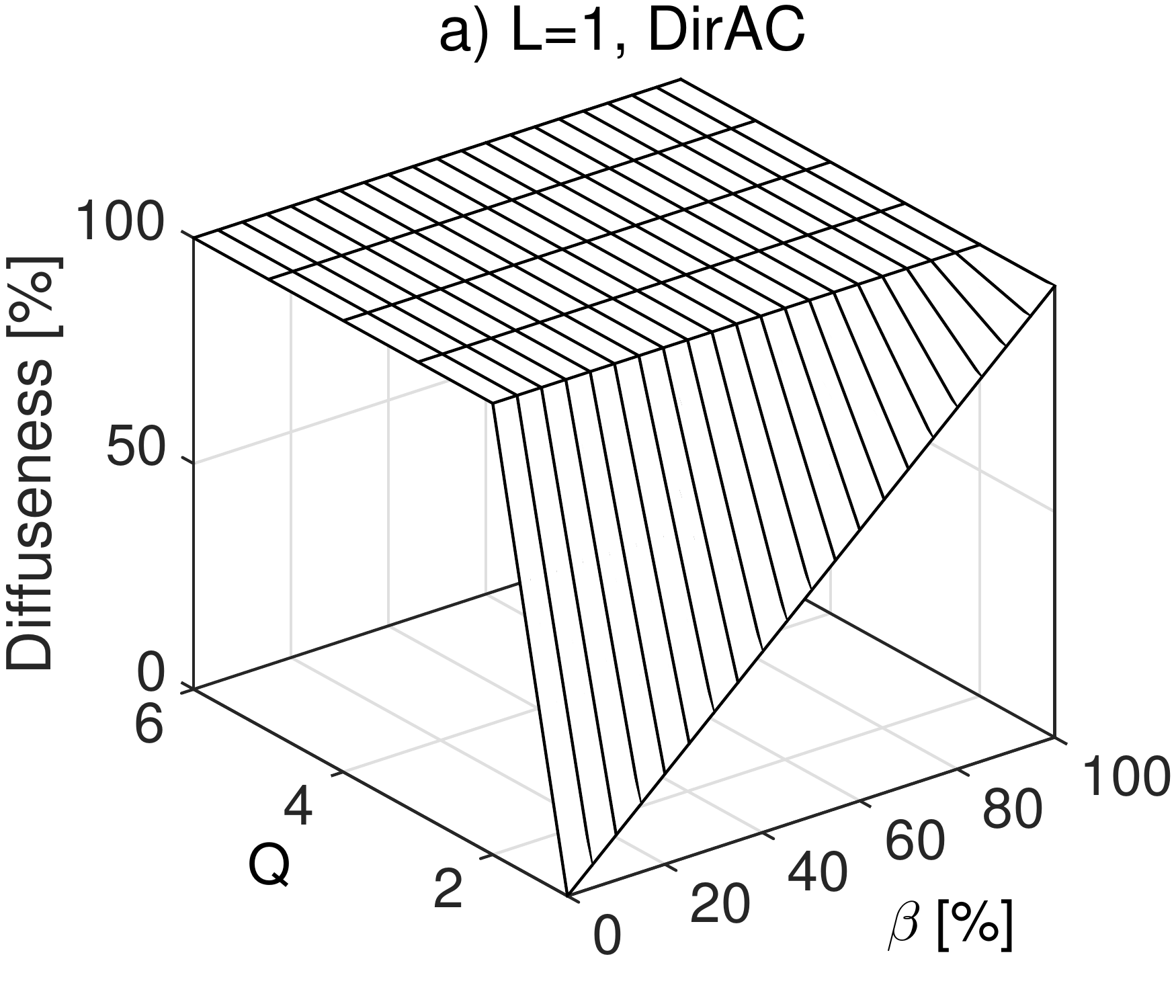}
}
\centerline{
\includegraphics[width=.48\columnwidth]{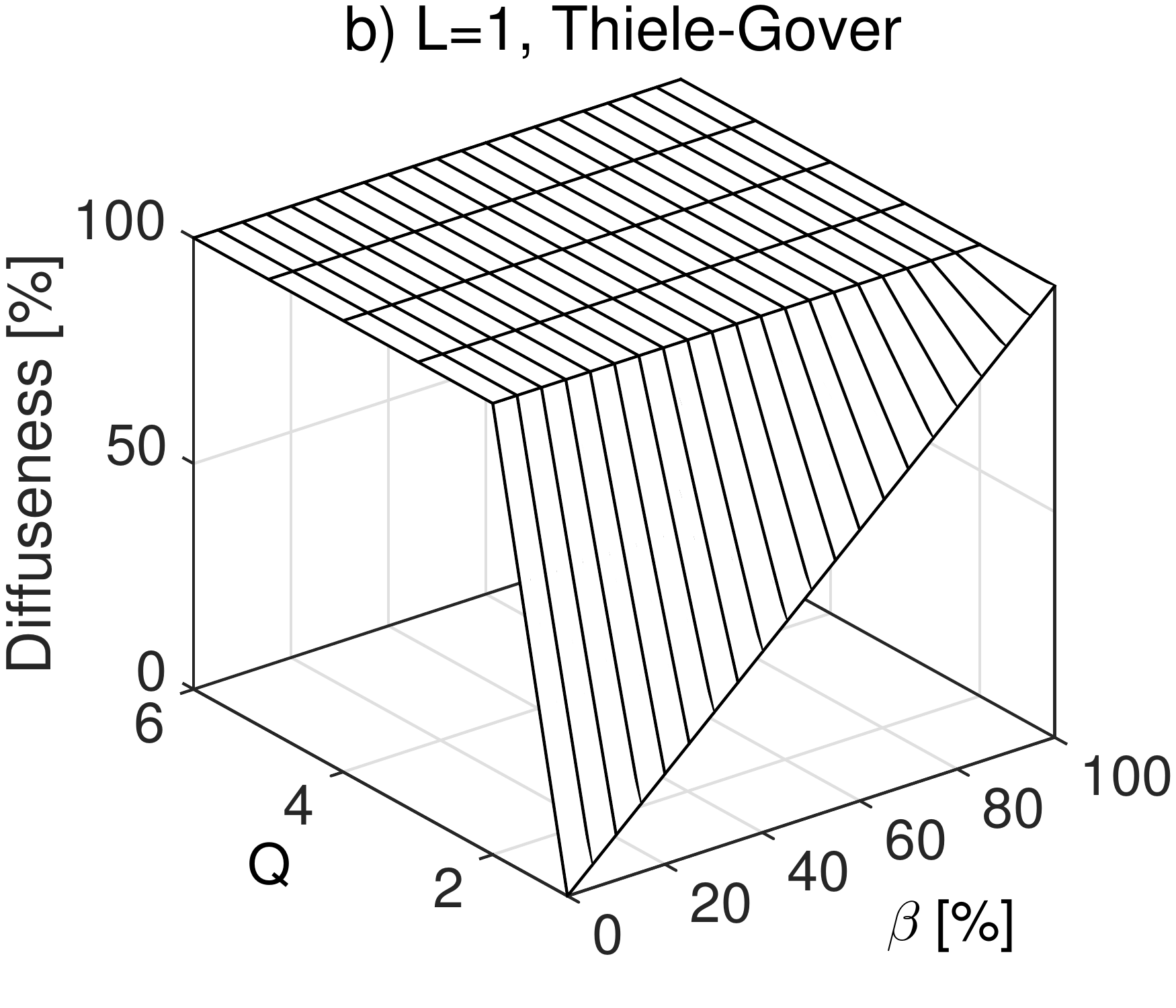}
\includegraphics[width=.48\columnwidth]{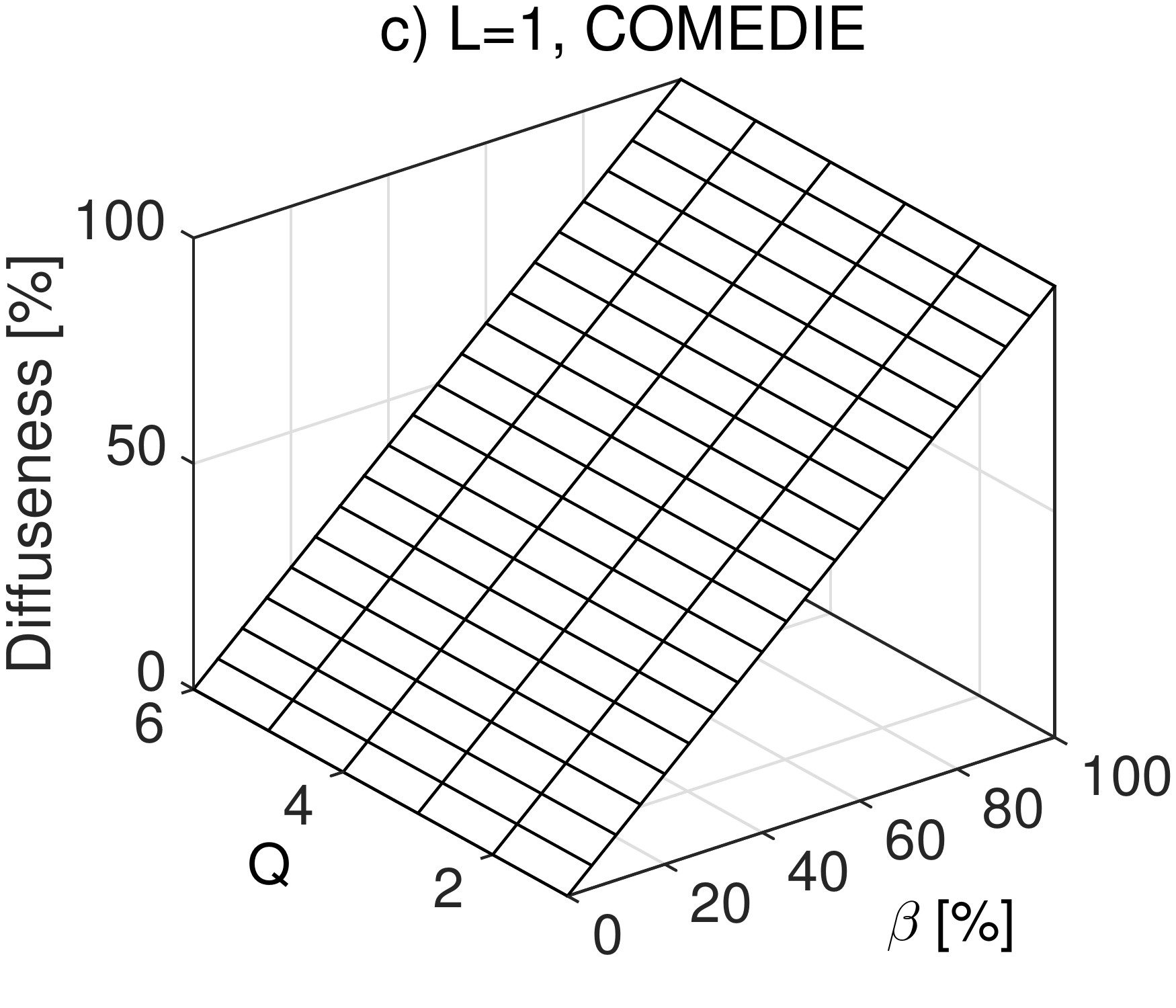}
} 
\centerline{
\includegraphics[width=.48\columnwidth]{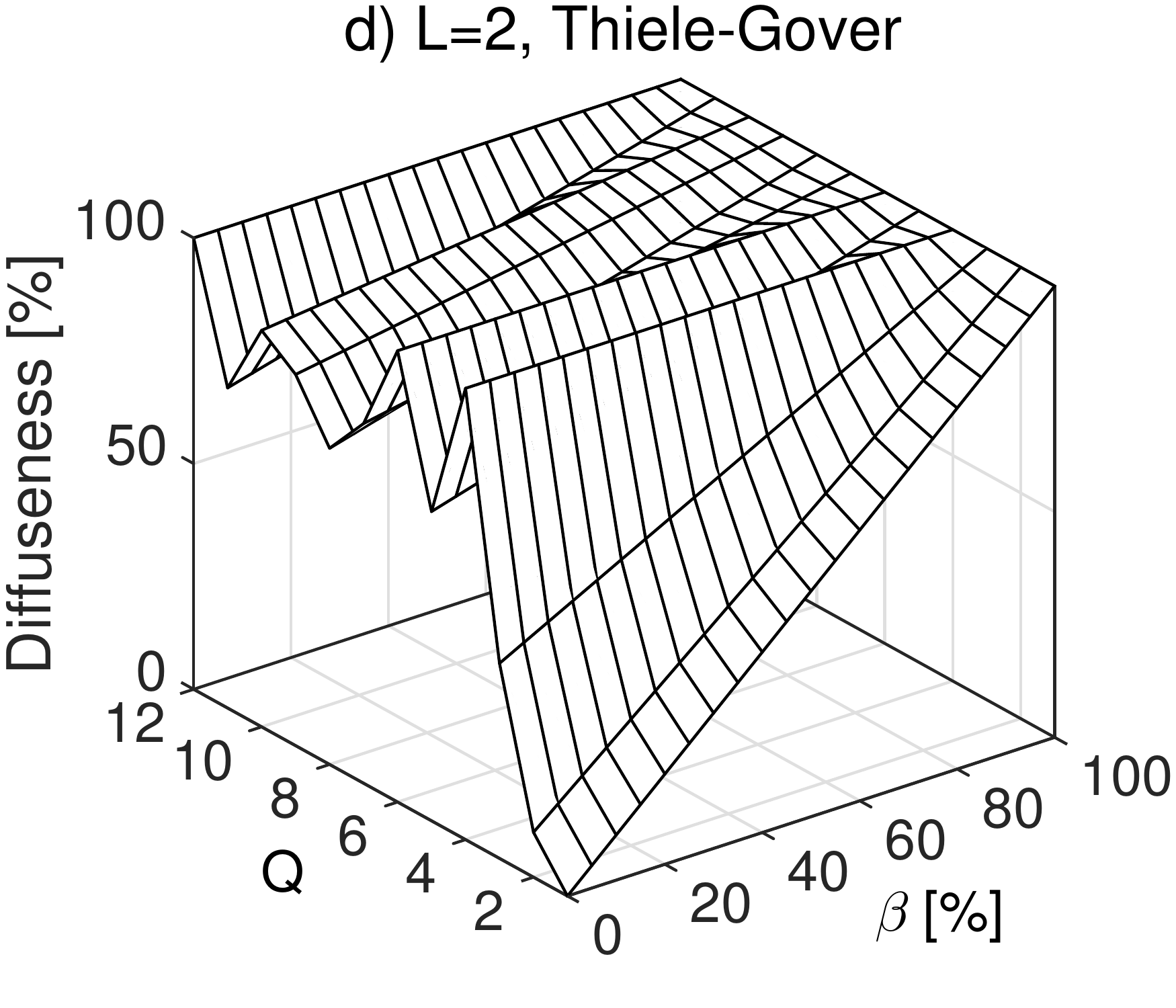}
\includegraphics[width=.48\columnwidth]{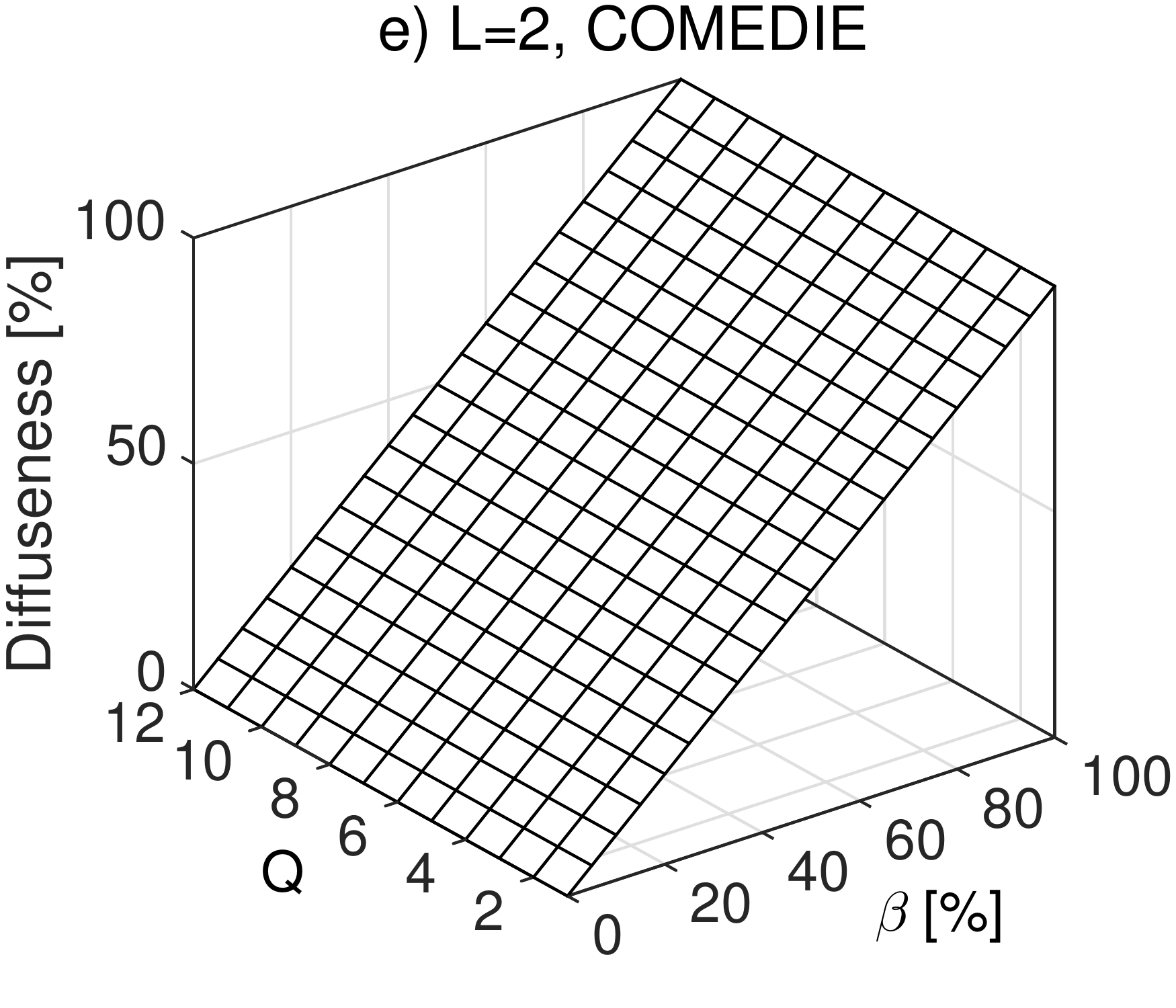}
} 
\centerline{
\includegraphics[width=.48\columnwidth]{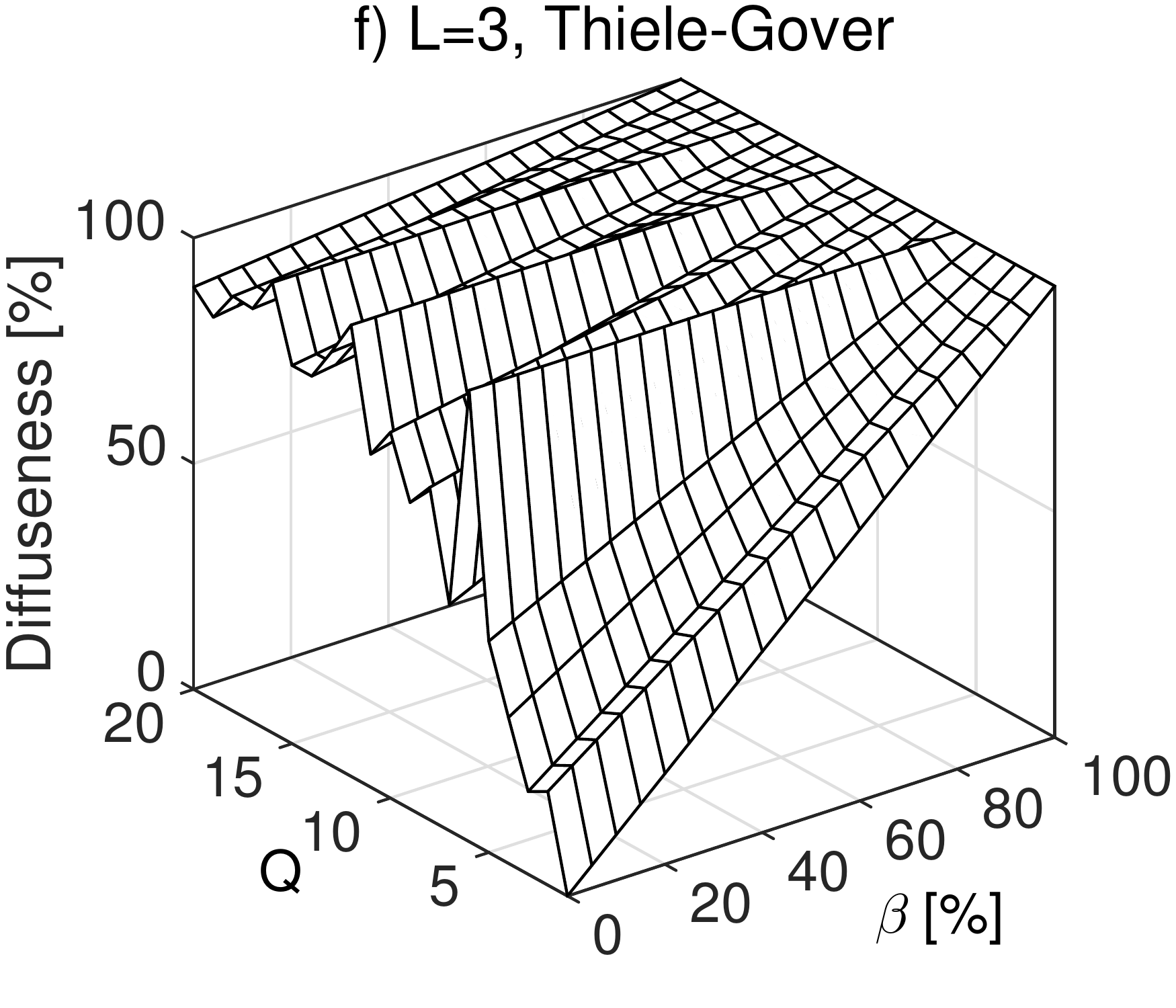}
\includegraphics[width=.48\columnwidth]{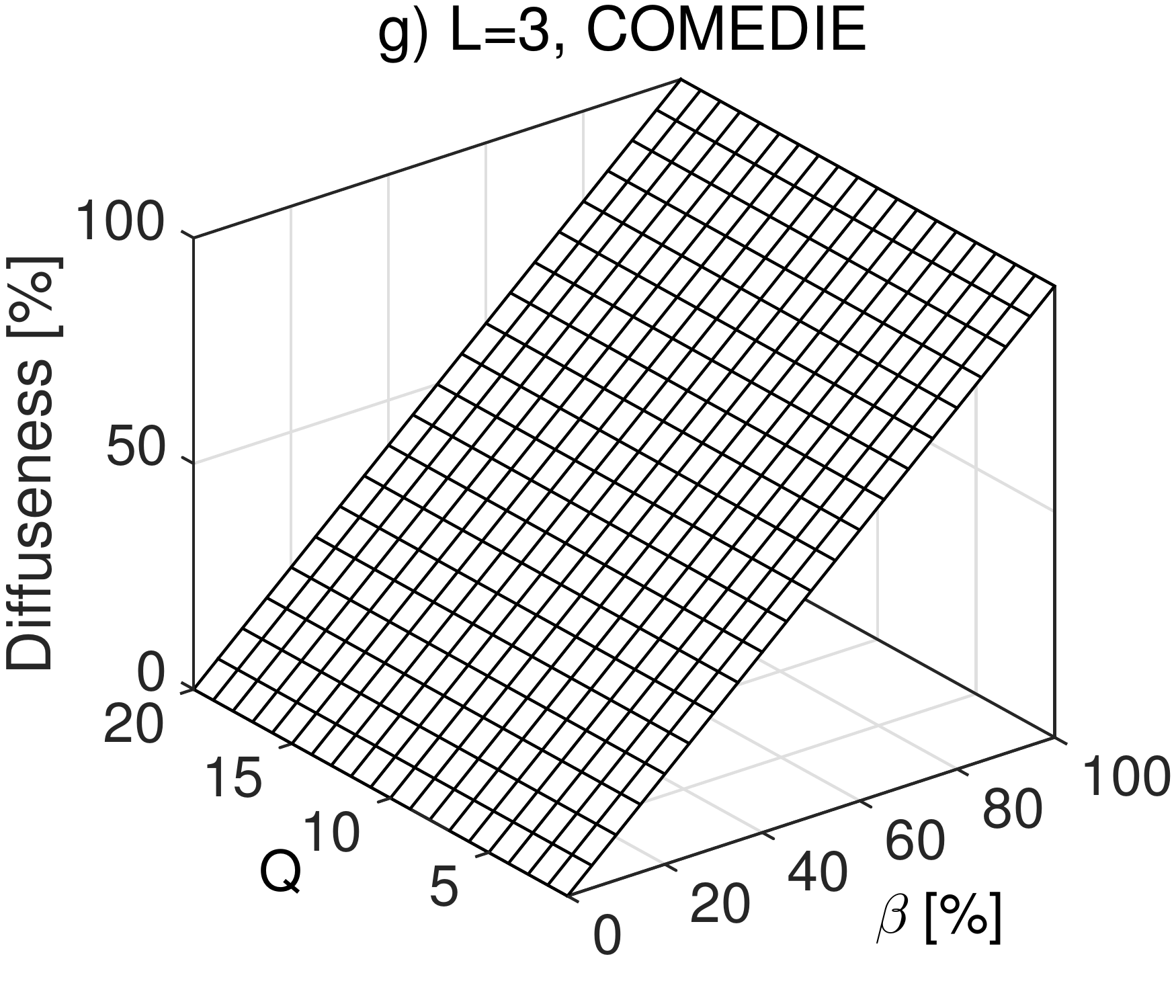}
} 
%\vspace{2mm}
%\centerline{\includegraphics[width=.48\columnwidth]{./figures/colorbarHoriz.pdf}}
\caption{This figure shows the estimated diffuseness in the presence of perfectly correlated sound sources and a perfectly diffuse component. The plots represent the estimated diffuseness as a function of the number of sound sources and diffuse sound field relative energy, using: a) The DirAC method with order-1 SH signals; b-d-f) The Thiele-Gover method applied to SH signals with orders 1, 2 and 3; c-e-g) the COMEDIE method applied to SH signals with orders 1, 2 and 3.}
\label{fig:CompCorSou}
\end{figure}
 
The results are illustrated in Figure~\ref{fig:CompNumSou} for SH orders 1, 2 and~3. In this figure, diffuseness is plotted as a function of both~$Q$, the number of plane waves, and $\beta$, the relative noise level using a surface mesh plot. Diffuseness generally increases as either $Q$ or  $\beta$ increases, but it is the way in which diffuseness increases that is important here. In Figures~\ref{fig:CompNumSou}f~and~\ref{fig:CompNumSou}g,  we see a fairly smooth increase in diffuseness for the order-3 SH signals as a function of both $Q$ and $\beta$. As the order of the SH signals decreases, we observe that diffuseness increases more sharply as a function of $Q$. 

We now consider Figure~\ref{fig:CompNumSou} in more detail. To begin, examine the bottom-right edge of all of the surface meshes, which corresponds to $Q = 1$.  We see that, in the presence of only one plane-wave source and a perfectly diffuse noise component, all methods and SH orders perform equally well and provide diffuseness estimates that are approximately equal to the relative noise level, $\beta$. Examine now the results for the order-1 SH signals only (Figures~\ref{fig:CompNumSou}a,~\ref{fig:CompNumSou}b, and~\ref{fig:CompNumSou}c). The DirAC method provides a diffuseness estimate of 100\% for $Q=2$. As explained previously, the presence of at least two sound sources located in opposite directions and having equal powers makes the sound field look perfectly diffuse with the DirAC estimate. In contrast, we see that it takes four sources evenly distributed on the sphere ($Q = 4$) to make the sound field look perfectly diffuse with the Thiele-Gover and COMEDIE diffuseness estimates.

Considering the results obtained with order-2 and order-3 SH signals, we see that diffuseness increases more gradually with $Q$ as the SH order increases. Note that the Thiele-Gover and the COMEDIE estimates behave similarly: increasing the number of plane-wave sources makes the sound field appear more and more diffuse. This is as expected because increasing the number of sources makes the situation more similar to the ideal diffuse condition, where an infinite number of sources are evenly distributed in space. Significantly, there are two important differences between the results obtained with the Thiele-Gover and COMEDIE diffuseness estimates. First, the diffuseness estimate increases more slowly as a function of $Q$ for the COMEDIE algorithm compared to the Thiele-Gover algorithm. In other words, for a given $Q$ and $\beta$ value, the COMEDIE diffuseness estimate is closer to $\beta$ than the Thiele-Gover estimate. For example, at order~3 and when $\beta=0$, at least $17$~plane-wave sources are required for the estimated diffuseness to be greater than $90\%$ with the COMEDIE method, while only $14$~are required with the Thiele-Gover method (refer to the bottom-left edge of the meshes in Figures~\ref{fig:CompNumSou}f and \ref{fig:CompNumSou}g). Second, the COMEDIE diffuseness increases with $Q$ more \emph{linearly} than the Thiele-Gover diffuseness. Examining the values obtained for order-3 signals with $\beta=0$, we see that it takes approximately 4, 8, 12 and 17 sources for the COMEDIE diffuseness estimate to reach 20, 50, 70 and $100\%$, respectively. In other words, the COMEDIE estimate for $\beta=0$ is roughly equal to $Q/(L+1)^2$. This stands to reason as the rank of the SH signal covariance matrix should be equal to the number of sources. On the other hand, the Thiele-Gover estimate increases very quickly for a small number of sources and then more slowly when there are more than 5 or so sources.

The numerical simulations were repeated with perfectly correlated sources, \emph{i.e.} having all of the plane-wave sources emitting the same 1024-sample long Gaussian white-noise signal. The other parameters, such as the source directions, were kept identical to that of the first scenario. Note that a large number of perfectly correlated sound sources does not create a diffuse sound field. The results of the simulations are presented in Figure~\ref{fig:CompCorSou}. The COMEDIE algorithm behaves differently from the two other methods. For orders~1,~2 and~3, the COMEDIE estimate is approximately equal to $\beta$, independent of the number of sources in the directional component. This is because for the COMEDIE algorithm, the presence of $Q$ perfectly correlated sources has an identical impact to that of only one source. In contrast, both the order-1 DirAC and Thiele-Gover estimates yield large diffuseness values when there are more than one source present. 

In summary, among the three considered methods, the COMEDIE diffuseness estimate seems to more adequately characterize 
both diffuseness arising from the presence of a diffuse noise background and diffuseness arising from the presence of multiple uncorrelated sources distributed in space. In particular, unlike the DirAC and Thiele-Gover estimates, the COMEDIE estimate is not affected by the presence of multiple correlated sources. However, because the order-$L$ COMEDIE diffuseness is a single number, it provides no means to discriminate between the two aforementioned types of diffuseness. This issue is addressed in the next section.

\section{Diffuseness Profiles}
\label{sec:profiles}
We now introduce the concept of diffuseness profiles. We have shown how the order-$L$ SH signals can be analyzed to estimate the diffuseness of a sound field. The numerical simulations presented in Section~\ref{subsec:Simul} show that the diffuseness estimates vary depending on the order $L$ of the SH signals. In this section, we show that the diffuseness values obtained for order 1, 2, ..., $L$ are not contradictory, but rather provide complementary and useful information regarding the spatial properties of the sound field.

\subsection{The Diffuseness Profile Concept}
We have seen in the previous section that, for a given scenario defined by a relative noise level $\beta$ and a number of plane-wave sources $Q$, the diffuseness values calculated using the SH signals up to order~1 may be different from the values obtained using the SH signals up to order~3. Specifically, observing Figure~\ref{fig:CompNumSou}, we see that for a given $\beta$ the order-1 estimate is always greater than the order-3 estimate. This arises because the SH signals up to order~3 provide higher spatial resolution. Intuitively, the sound field generated by a few uncorrelated sources evenly distributed in space may seem very diffuse when looking at only the order~1 SH signals, but the dominant sound propagation directions can be resolved by analyzing SH signals up to order~3. Thus, the order-3 diffuseness estimate is generally closer to $\beta$ than the order-1 estimate. Nevertheless, the order~1 estimate is not without benefits as we shall see.

\begin{figure}[!t]
\centerline{
\includegraphics[width=.3\columnwidth]{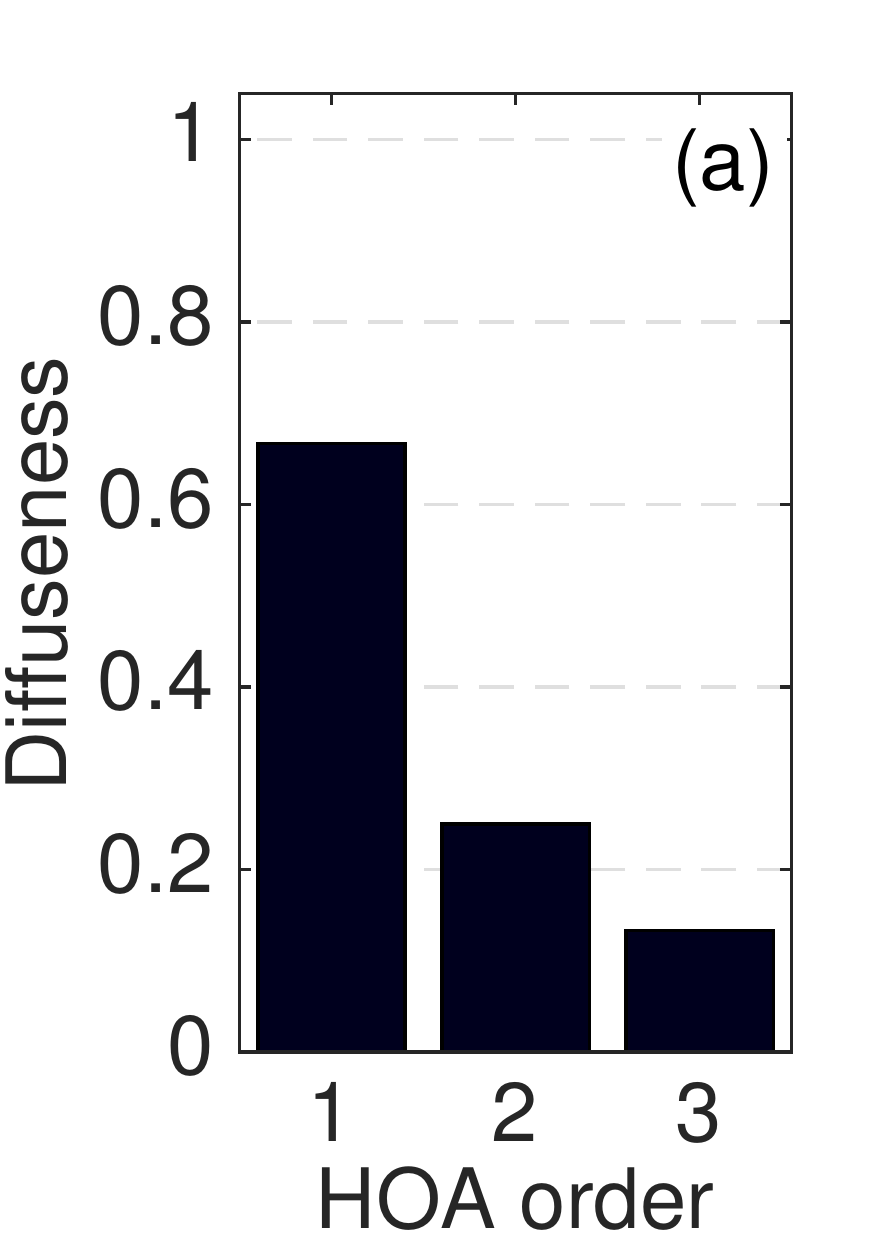}
\includegraphics[width=.3\columnwidth]{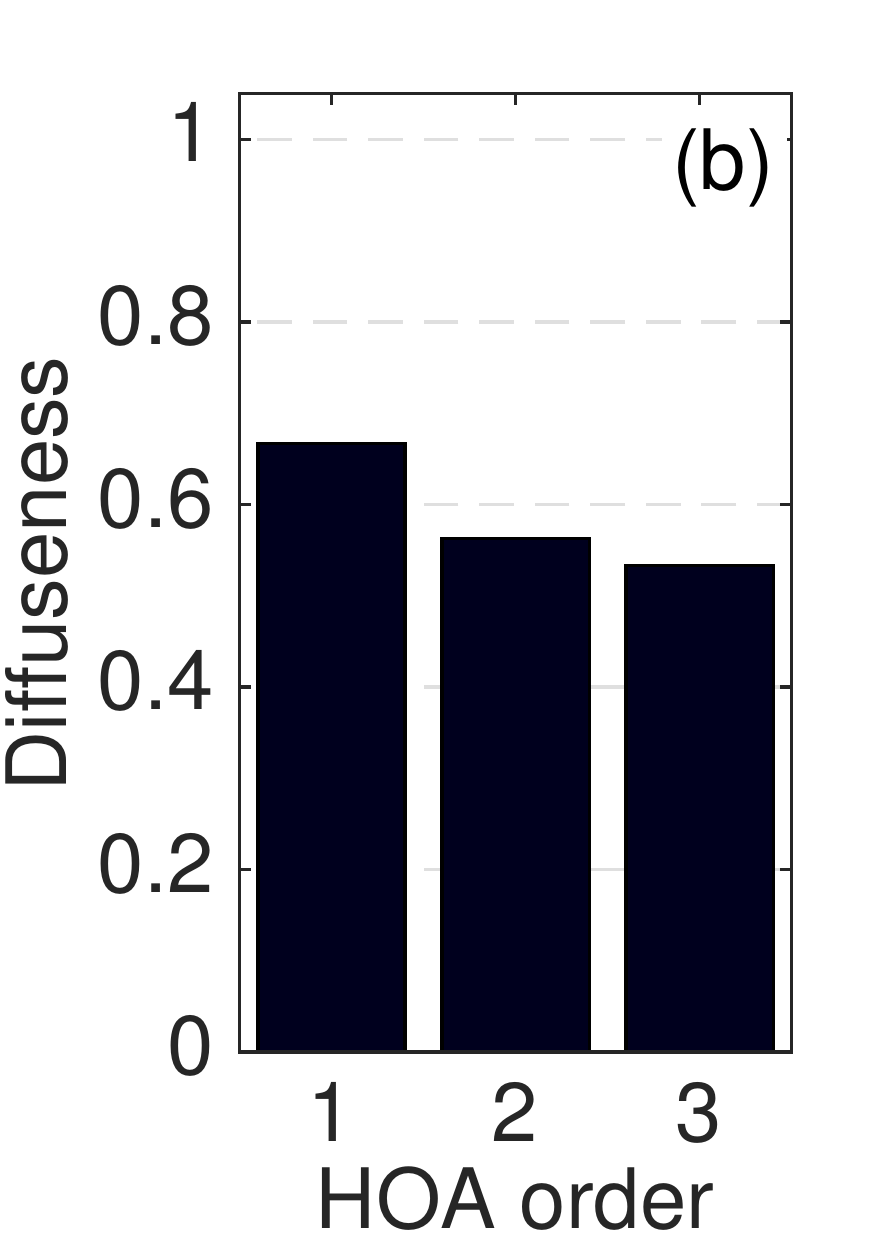}
\includegraphics[width=.3\columnwidth]{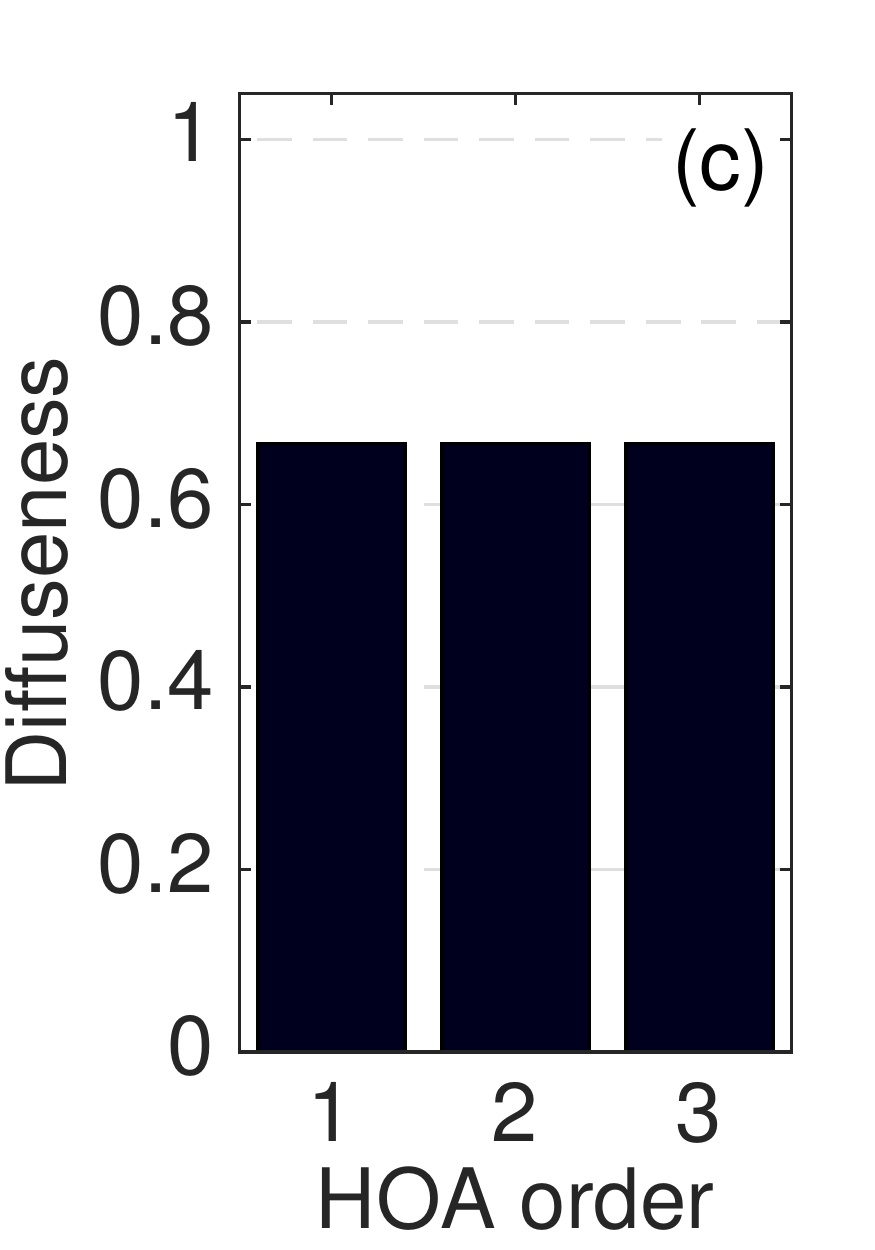}
}
\caption{Order-3 diffuseness profiles obtained in the presence of: a) 3 plane waves with $\beta=0$ ; b) 2 plane waves with $\beta=0.5$ ; c) 1 plane waves with $\beta=\frac{2}{3}$.}
\label{fig:DiffProfOrd1}
\end{figure}

\begin{figure}[!t]
\centerline{
\includegraphics[width=.3\columnwidth]{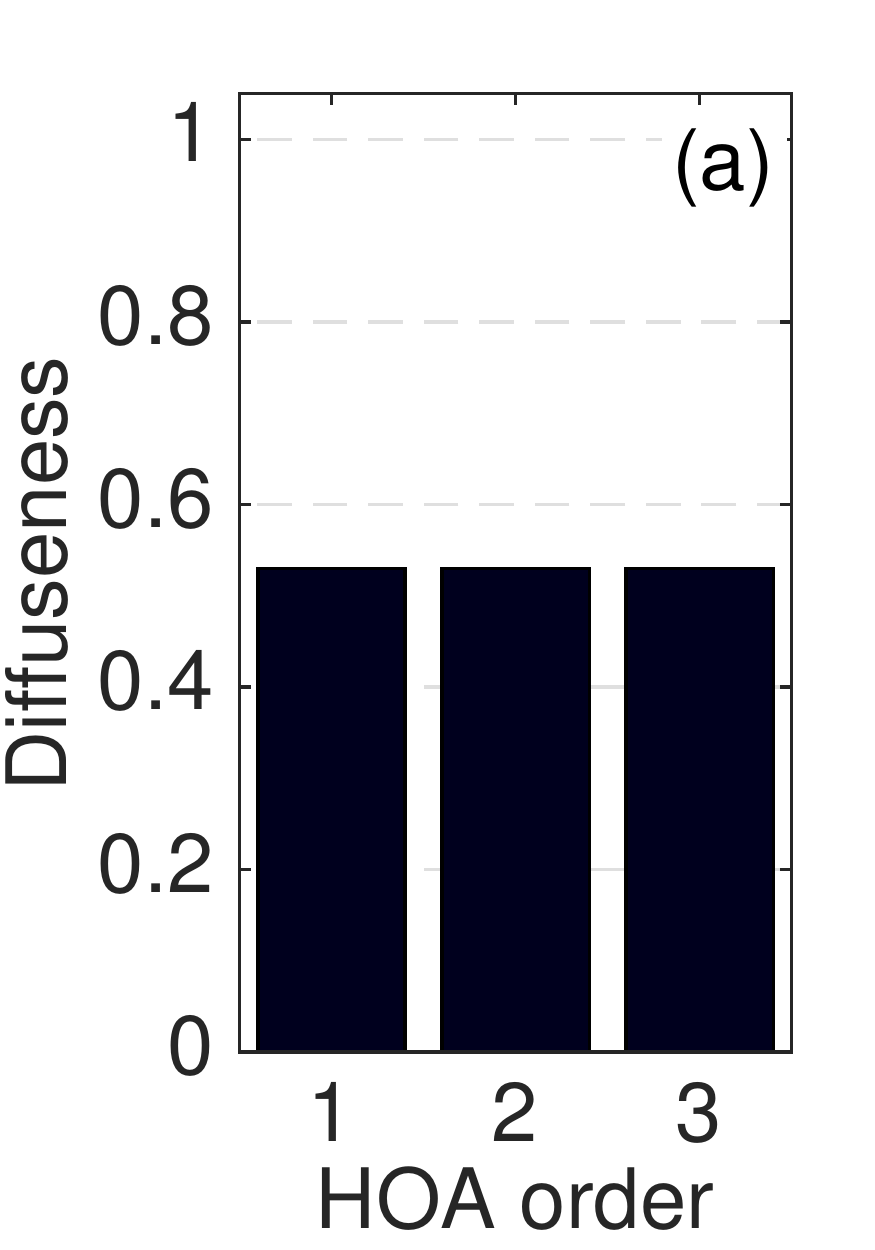}
\includegraphics[width=.3\columnwidth]{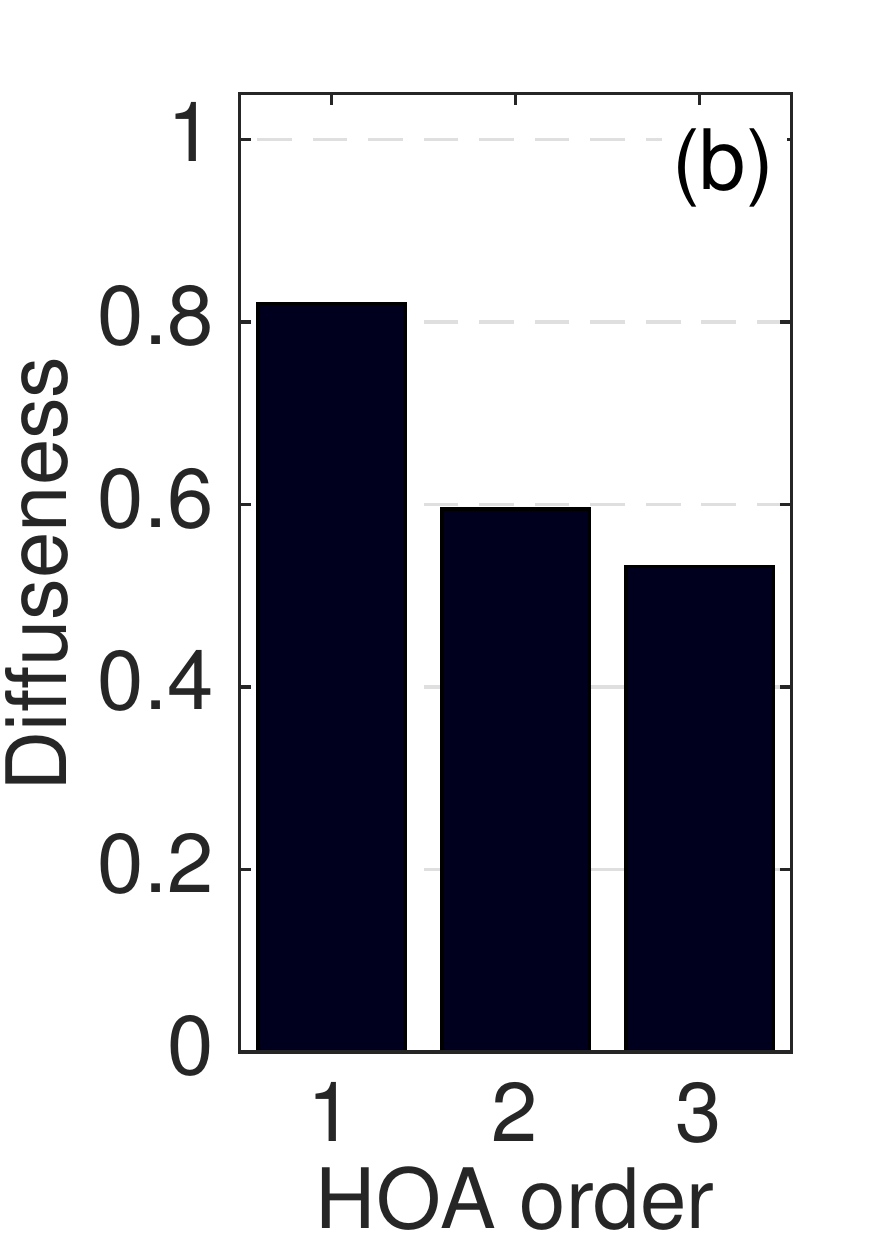}
\includegraphics[width=.3\columnwidth]{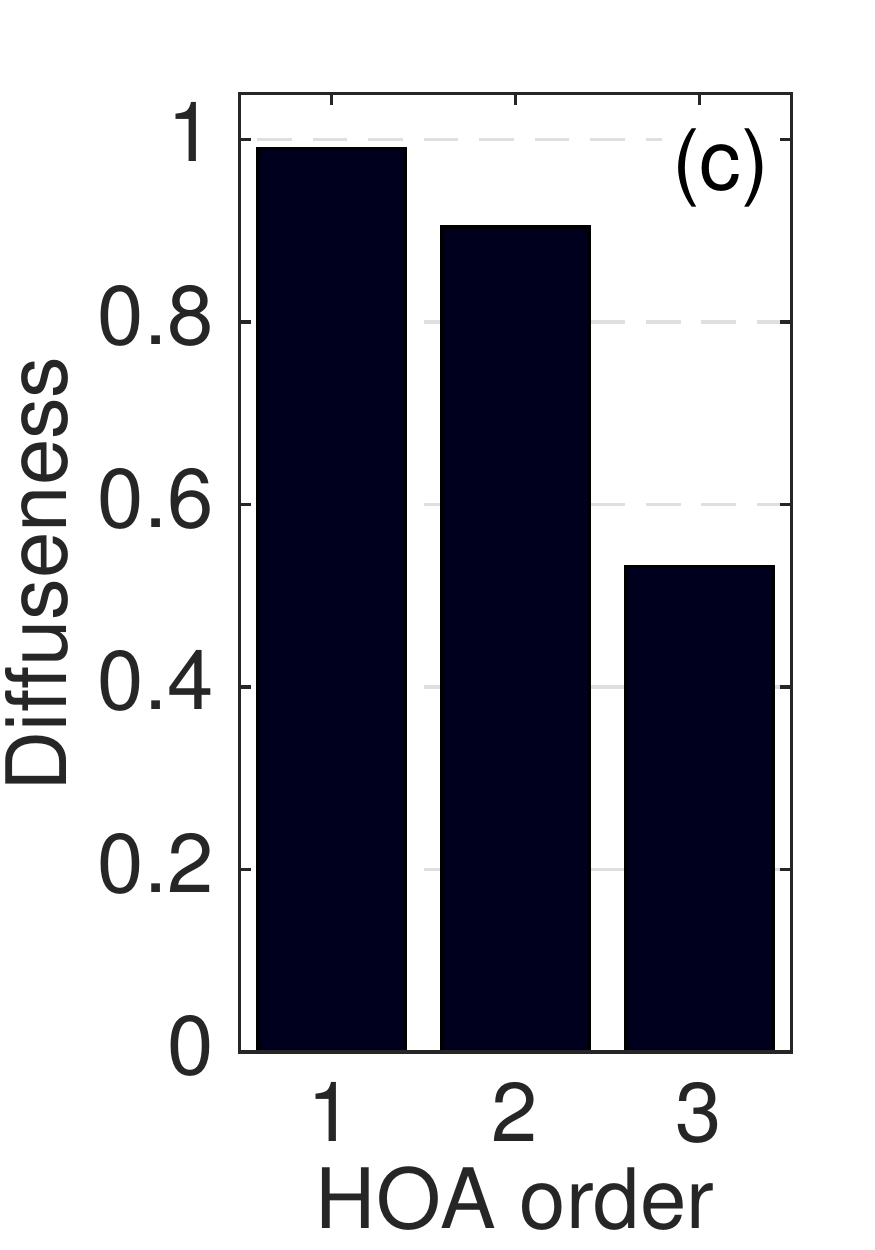}
}
\caption{Order-3 diffuseness profiles obtained in the presence of: a) 1 plane wave with $\beta=0.53$ ; b) 3 plane waves with $\beta=0.46$ ; c) 9 plane waves with $\beta=0$.}
\label{fig:DiffProfOrd3}
\end{figure}

A critical observation is that different sound field scenarios may yield the same diffuseness estimate for a given SH order, but vary in other SH orders. This indicates there is likely value in considering the diffuseness estimates across SH orders. To begin, consider Figure~\ref{fig:DiffProfOrd1} where the values of the order-1, 2 and 3 COMEDIE diffuseness estimates are plotted for three different sound field configurations. Note that the SH signals were generated as described in Section~\ref{subsec:Simul}. The order-1 diffuseness is equal to $2/3$ for the three different scenarios, while the order-2 and order-3 diffuseness estimates vary. Significantly, a similar observation can be made regarding the order-3 diffuseness estimate, as illustrated in Figure~\ref{fig:DiffProfOrd3}. In this case, the order-3 diffuseness estimate is always equal to $0.53$, while the order-1 and order-2 diffuseness estimates vary across scenarios. In other words, although the order-3 diffuseness estimate is generally closer to $\beta$, values obtained across the various SH orders provide complementary information regarding the sound field.

Based on the observations above, we suggest that when characterizing the spatial properties of a sound field, it is useful to observe the order-1, 2, ..., $L$ diffuseness estimates \emph{together}, which we refer to as observing the order-$L$ diffuseness profile. Examining Figures~\ref{fig:DiffProfOrd1} and~\ref{fig:DiffProfOrd3}, we can make the following observations regarding the interpretation of diffuseness profiles. A flat profile generally indicates that there is at most one dominant sound propagation direction, which can be caused by the presence of one source or several closely located sources. Another possible interpretation is that the sound sources are strongly correlated with each other. On the other hand, a profile with a decreasing slope indicates that several uncorrelated sources are located in directions opposite to each other, which makes the sound field look more diffuse when looking only at the lower order SH signals. In summary, diffuseness profiles help determine whether diffuseness arises from the presence of a diffuse noise background or from the presence of multiple yet countable uncorrelated sources distributed in space.

\subsection{Measurement of Room Acoustics Using an SMA} 
\label{sec:expe}
In order to illustrate the diffuseness profile applied to a real acoustic scenario, we now present the results of diffuseness profile measurements made in a room using an SMA.

\begin{figure}[t]
\centerline{\includegraphics[width=\columnwidth]{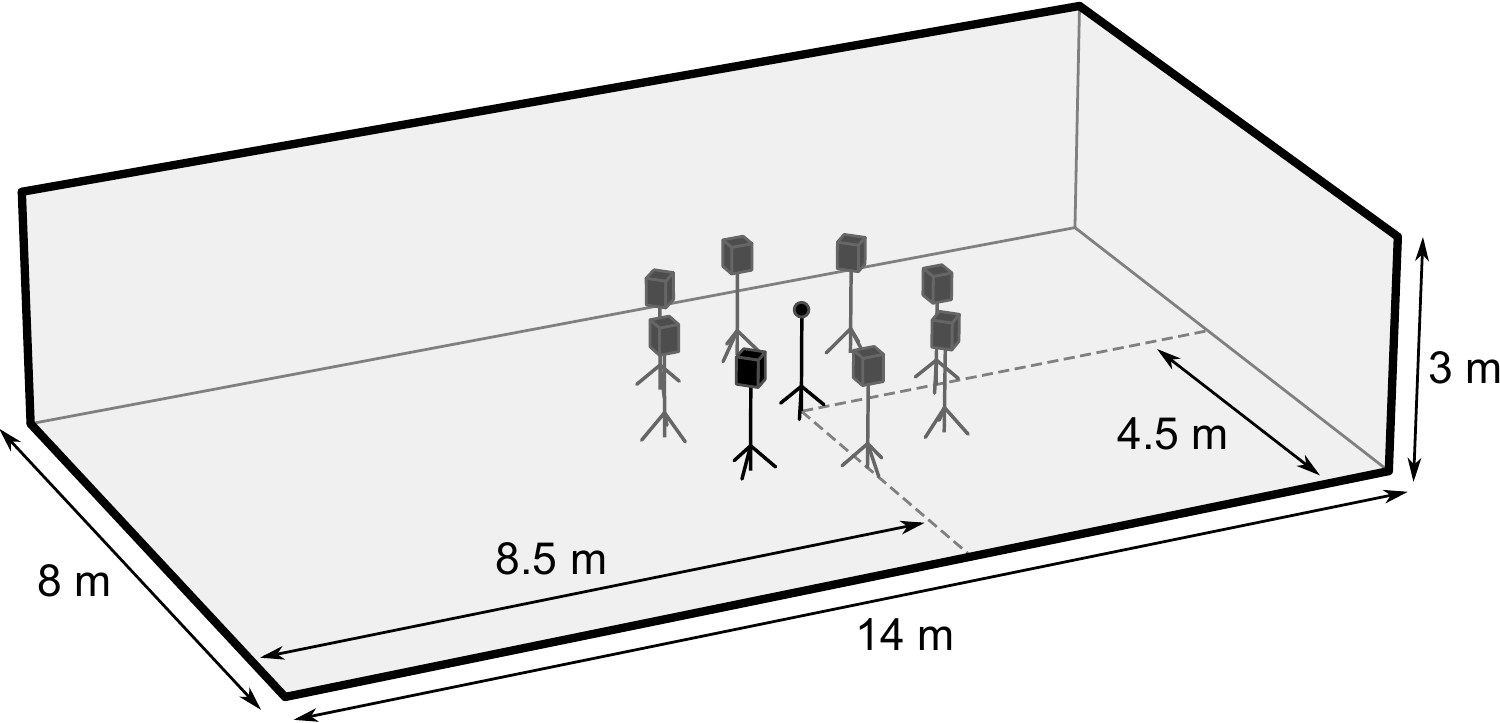}}
\caption{Illustration of the measurement setup for the experiment described in Section~\ref{sec:expe}}
\label{fig:MeasPhoto}
\end{figure}

A diagram illustrating the measurement setup is shown in Figure~\ref{fig:MeasPhoto}. The microphone array was a dual-concentric SMA consisting of 64 microphones distributed on the surface of two spheres: 32 microphones were located on a rigid sphere of radius 16.3 mm, and the 32 others were located on an open sphere of radius 60 mm. Impulse responses were measured using a Tannoy V6 loudspeaker that was moved to 8 different positions around the SMA. The distance between the loudspeaker and the center of the SMA was 1.8~m for the eight speaker positions, while the azimuthal angle varied with~$45^\circ$ steps. Note that the loudspeaker was positioned using a tape measure and simple geometry principles, which resulted in a placement accuracy on the order of a few centimeters. Therefore the measured impulse responses were resynchronized so that the signals emitted by the speaker located at the different positions were perfectly in phase at the center of the SMA. The room in which the measurements took place is an office space with dimensions $14\times8\times3$~m. Its T60 reverberation time is 468~ms and the average direct-to-reverberant ratio (DRR) for the eight speaker positions was 4.5~dB.

The measured impulse responses were used to generate microphone signals corresponding to various scenarios. The microphone signals were then filtered to obtain order-3 SH signals using the process described in~\cite{Jin2013}. The SH signals were band-pass filtered between 1.6~kHz and 16~kHz, which corresponds to the frequency range over which this SMA provides order-3 SH signals with only a moderate amount of noise (SNR $\ge$ 15~dB). Lastly, the order-3 diffuseness profiles were calculated using the COMEDIE algorithm.

In the first scenario we simulated the presence of multiple sound sources emitting perfectly uncorrelated signals in the room. The microphone signals corresponding to the different sources were calculated individually by convolving the measured impulse responses with uncorrelated, Gaussian, white noise signals with equal power, and then summed together. Four different configurations with different azimuthal angles were simulated: 1 source ($0^\circ$), 2 sources ($0^\circ$ and $180^\circ$), 4 sources ($0^\circ, 90^\circ, 180^\circ$ and $270^\circ$) and 8 sources ($0^\circ, 45^\circ, 90^\circ, \dots, 315^\circ$).

\begin{figure}[t]
\centerline{\includegraphics[width=\columnwidth]{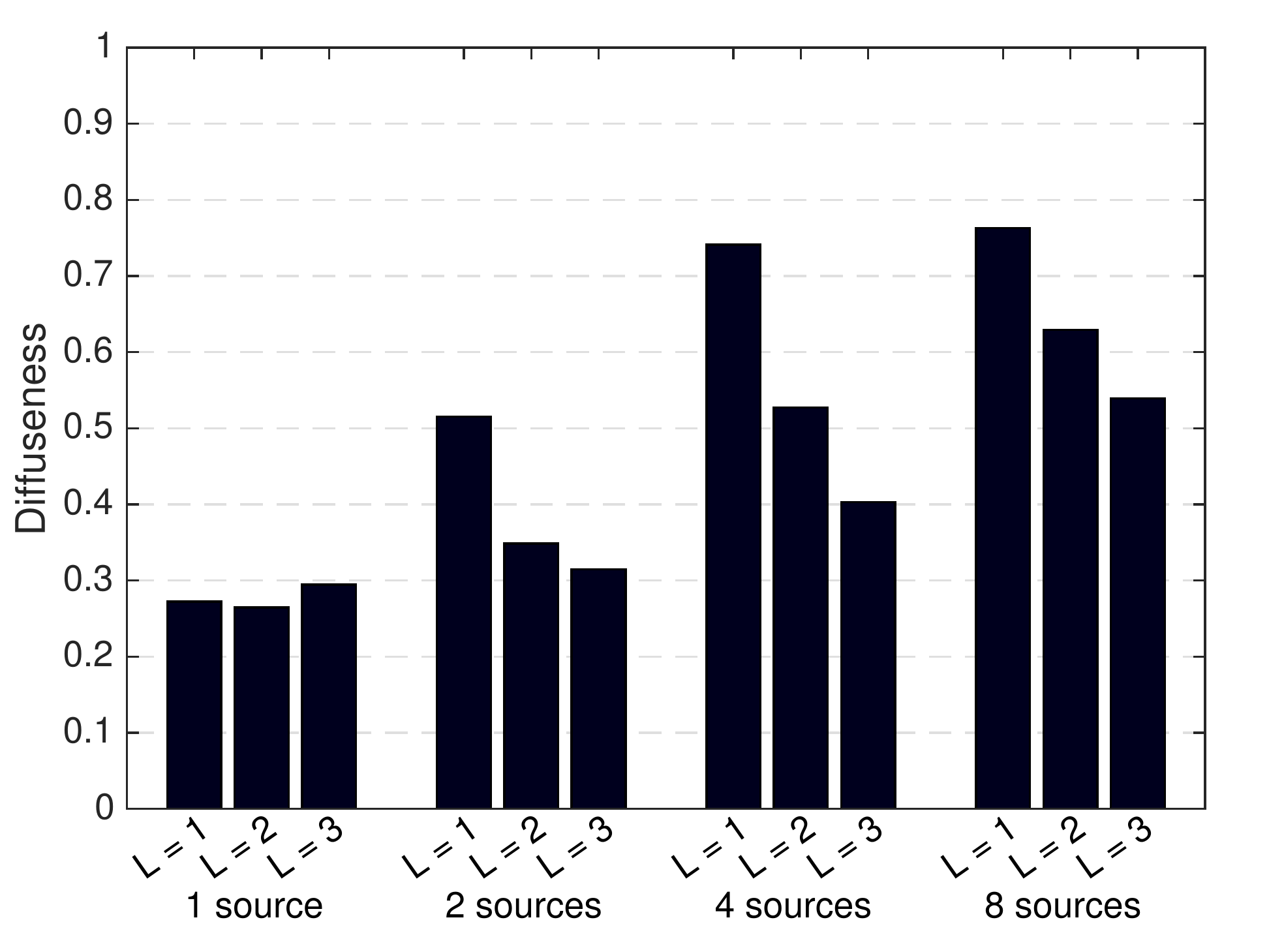}}
\caption{Measured order-3 diffuseness profiles in the presence of perfectly uncorrelated sources in a room.}
\label{fig:MeasProfCorr}
\end{figure}

The corresponding diffuseness profiles are illustrated in Figure~\ref{fig:MeasProfCorr}. When only one source is present the diffuseness profile is almost flat as expected. In this scenario the estimated diffuseness value for orders~1,~2 and~3 is approximately 0.26,~0.26 and 0.29, respectively. These values are well in line with the measured DRR. Assuming the reverberated part of the sound field is perfectly diffuse and uncorrelated with the direct part, the DRR is equal to the direct-to-diffuse ratio (DDR), which is the ratio of the direct energy to the diffuse energy. The relative noise level is thus given by~\cite{DelGaldo2012}:
\begin{align}
\beta=\frac{1}{1+\mathrm{DDR}} =\frac{1}{1+10^{\frac{4.5}{10}}}\approx0.26 \ \text{.}
\end{align}
The fact that the estimated order-3 diffuseness is slightly larger than this value can be explained by the fact that the order-3 SH signals are slightly noisier than the order~1 and order~2 SH signals, despite the band-pass filtering. Errors in the SH signals may occur, for instance, due to microphone misplacement and miscalibration. The presence of noise is then interpreted by the COMEDIE algorithm as additional diffuse energy.

Regarding the scenarios with 2,~4 and~8 sources, we observe that when the number of sources increases, the \mbox{order-1} diffuseness increases quickly, whereas the order-2 and~3 diffuseness increase more slowly. This occurs because, as shown in the simulation results in Section~\ref{subsec:Simul}, the higher the SH order, the less sensitive the COMEDIE estimate is to the presence of discrete sources located in opposite directions. Lastly, note that the \mbox{order-1} diffuseness obtained in the presence of eight sources is almost identical to that obtained in the presence of four sources. This arises because, from the point of view of the order-1 SH signals, both situations are equivalent to having sound sources emitting uncorrelated signals in every direction in the horizontal plane. Hence adding more sources in the horizontal plane does not increase the order-1 diffuseness. A larger diffuseness value could only be obtained if a source was located at a higher or lower elevation, as this would make the source distribution slightly closer to the perfectly diffuse scenario in which sources are evenly distributed over the sphere.

\begin{figure}[t]
\centerline{\includegraphics[width=\columnwidth]{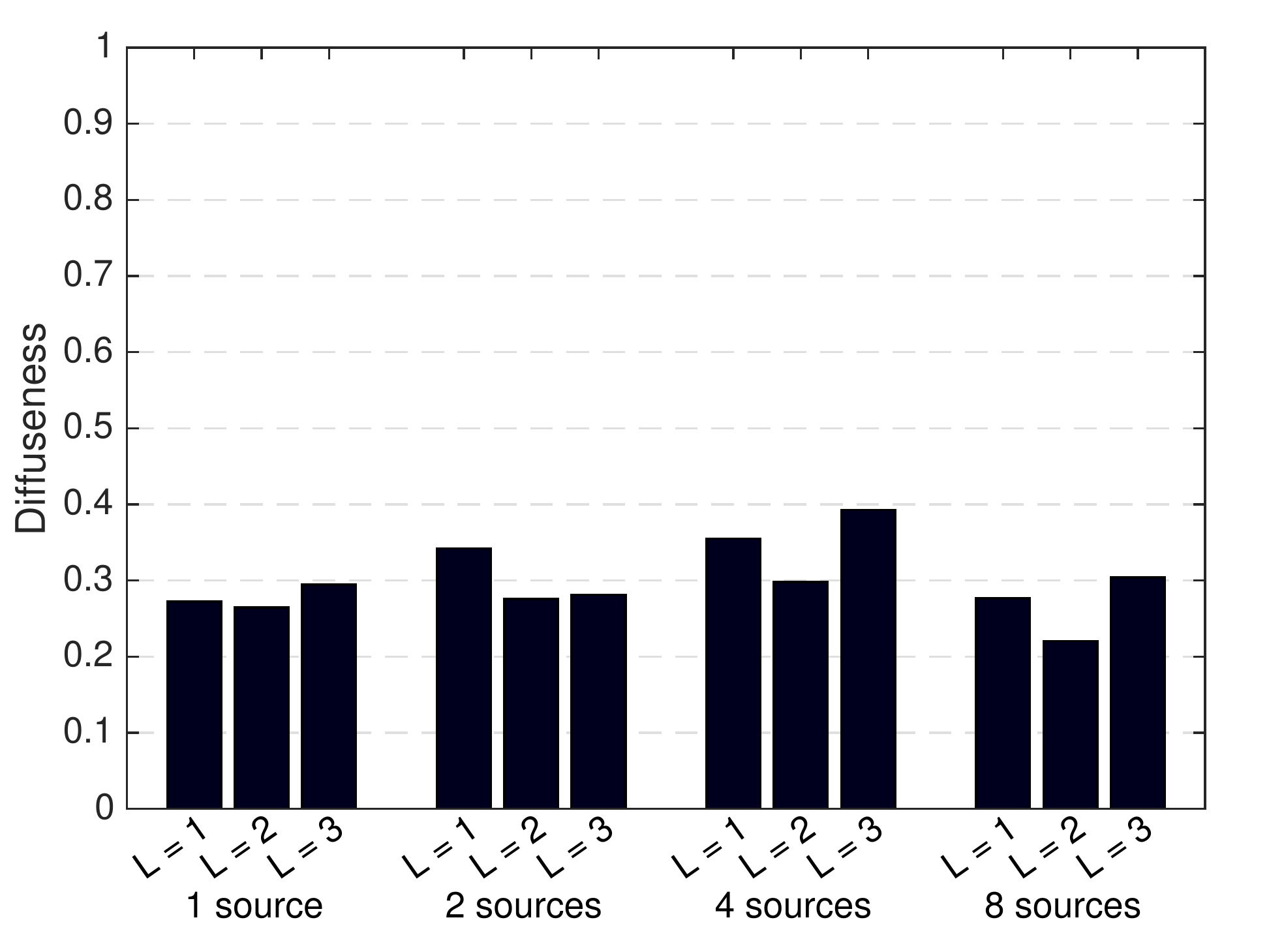}}
\caption{Measured order-3 diffuseness profiles in the presence of perfectly correlated sources in a room.}
\label{fig:MeasProfUncorr}
\end{figure}

In a second scenario, we analyzed the presence of multiple sources emitting perfectly correlated signals in the room. The SH signals were generated using the same procedure as in the first scenario, with the exception that the same white noise signal was used for every source position. The measured diffuseness profiles are illustrated in Figure~\ref{fig:MeasProfUncorr}. The diffuseness profiles remain almost identical as the number of sources increases. This occurs because the sources are perfectly correlated and thus result in only one large eigenvalue in the SH signal covariance matrix. Therefore, the estimated diffuseness values obtained with two, four and eight correlated sources are on the order of that obtained with only one source. In fact, the diffuseness even decreases slightly in the case where there are eight sources. This arises due to the positive interference between the sources, as the energy of the direct sound field increases relative to the energy of the reverberated sound field. Note that, if the Thiele-Gover diffuseness estimate had been used, the plot corresponding to the second scenario would have been similar to that obtained for the first scenario.

\section{Conclusions}
\label{sec:concl}
In this paper we have investigated the characterization of sound field diffuseness using SMAs. In general terms, measuring diffuseness consists in measuring how similar the sound field is to the ideal, perfectly diffuse field. We have shown that two factors may determine diffuseness: 1) the presence of a diffuse noise background; and 2) the presence of multiple, uncorrelated sound sources distributed across space. We have proposed a new diffuseness estimate, the COMEDIE diffuseness estimate, which relies on the analysis of the spectrum of the SH signal covariance matrix. The COMEDIE diffuseness estimate generalizes existing diffuseness measures by taking into account the correlation of signals incoming from different directions. As well, we have introduced the concept of diffuseness profiles, which consist in observing the diffuseness at several SH orders simultaneously. Experimental results indicate that diffuseness profiles help characterize the nature of sound field diffuseness and assist with the disambiguation of diffuseness arising from multiple, uncorrelated sound sources compared with a diffuse noise background. 

\bibliographystyle{IEEETran}
\bibliography{IEEETransDiffusivity}

%\begin{biography}[{\includegraphics[width=1in,height=1.25in,clip,keepaspectratio]{mshell}}]{Michael Shell}

%\end{IEEEbiography}


%\newpage



\end{document}